%% file: main.tex
\documentclass[10pt,a4paper]{article}
\usepackage{amsmath, amssymb, graphics}
\usepackage{fullpage}
\usepackage{graphicx}

\newcommand{\Prob}[2]{\mathbf{P}_{#1} \left( #2 \right)}
\newcommand{\Expec}[2]{\mathbf{E}_{#1} \left[ #2 \right]}

\newcommand{\rbrdyn}{noisy best-response dynamics}
\newcommand{\ck}{{\sf CK}}
\newcommand{\welfare}{{\sf SW}}
\newcommand{\ignore}[1]{}
\renewcommand{\leq}{\leqslant}
\renewcommand{\geq}{\geqslant}

\newcommand{\proof}{\textit{Proof. }}
\newcommand{\qed}{\hspace{\stretch{1}$\square$}}

\newtheorem{definition}{Definition}

\newtheorem{lemma}[definition]{Lemma}

\newtheorem{theorem}[definition]{Theorem}
\newtheorem{cor}[definition]{Corollary}
\newtheorem{obs}[definition]{Observation}

\input{macro}

\begin{document}

\title{\bf Mixing Time and Stationary Expected Social Welfare of Logit Dynamics\thanks{Work partially supported by the PRIN 2008 research project COGENT (COmputational and GamE-theoretic aspects of uncoordinated NeTworks), funded by the Italian Ministry of University and Research.}}
\author{Vincenzo Auletta, Diodato Ferraioli, Francesco Pasquale, and Giuseppe Persiano \\[3mm]
Dipartimento di Informatica ``Renato M. Capocelli''\\
Universit\`a di Salerno \\ \texttt{auletta},\texttt{ferraioli},\texttt{pasquale},\texttt{giuper}@\texttt{dia.unisa.it}
}
\maketitle

\begin{abstract}
\input{./abstract}

\end{abstract}

\section{Introduction}
\input{./trunk/intro.tex}

\section{Markov chains summary and notation}\label{apx:markov}
\input{./trunk/Markov}

\section{The model and the problem}\label{sec::model}
\input{./trunk/model.tex}
\input{./trunk/coupling.tex}

\section{A $3$-player congestion game}\label{sec::ckgame}
\input{./trunk/ckgame.tex}

\section{Two player games}\label{sec::twoplayers}
\input{./trunk/twoplayers.tex}

\section{The OR game}\label{sec::orgame}
\input{./trunk/orgame.tex}

\subsection{Technical lemmas}\label{subsec:tech}
\input{./trunk/orgame-lemma}

\section{The XOR game}\label{sec:xorgame}
\input{./trunk/xor}

\section{Conclusions and open problems}\label{sec::conclusions}
\input{./trunk/conclusions.tex}

\bibliographystyle{plain}
\bibliography{logit}

\ignore{
\newpage
\begin{center}
\begin{LARGE}
\textbf{Appendix}
\end{LARGE}
\end{center}
\appendix

\input{./trunk/appendix.tex}

}

\end{document}

%% file: macro.tex
\newcommand{\x}{{\mathbf{x}}}
\newcommand{\y}{{\mathbf{y}}}
\newcommand{\hamming}{H}   %%macro Hamming distance
\newcommand{\ml}{{\nu_{\ell}}}
\newcommand{\Ml}{{\mu_{\ell}}}
\newcommand{\tm}{{t_\text{mix}}}
\newcommand{\tc}{{\tau_\text{couple}}}

%% file: abstract.tex
We study \emph{logit dynamics}~\cite{blumeGEB93} for strategic games. This dynamics works as follows: at every stage of the game a player is selected uniformly at random and she plays according to a \emph{noisy} best-response where the noise level is tuned by a parameter $\beta$. Such a dynamics defines a family of ergodic Markov chains, indexed by $\beta$, over the set of strategy profiles. We believe that the stationary distribution of these Markov chains gives a meaningful description of the long-term behavior for systems whose agents are not completely rational.

%of games where players are not fully rational: thus the stationary distribution can be used as a solution concept for such games.

Our aim is twofold: On the one hand, we are interested in evaluating the performance of the game at equilibrium, i.e. the expected social welfare when the strategy profiles are random according to the stationary distribution. On the other hand, we want to estimate how long it takes, for a system starting at an arbitrary profile and running the logit dynamics, to get close to its stationary distribution; i.e., the \emph{mixing time} of the chain.

In this paper we study the stationary expected social welfare for the $3$-player \ck\ game~\cite{ckSTOC05}, for $2$-player coordination games, and for two simple $n$-player games. For all these games, we also give almost tight upper and lower bounds on the mixing time of logit dynamics. Our results show two different behaviors: in some games the mixing time depends exponentially on $\beta$, while for other games it can be upper bounded by a function independent of $\beta$.

%% file: trunk/intro.tex
The evolution of a system is determined by its dynamics and 
complex systems are often described by looking at the
equilibrium states induced by their dynamics.
Once the system reaches an equilibrium state it stays there, 
thus equilibrium states 
describe the long-term behavior of the system.
In this paper we are mainly interested in systems whose individual components are \emph{selfish} agents.
The state of a selfish system is fully described by a vector of 
{\em strategies}, each controlled by one agent, 
and each state assigns a payoff to each agent.  
The agents are selfish in the sense that they 
pick their strategy so to maximize their payoff, given 
the strategies of the other agents.
\emph{Nash equilibrium} is the classical notion of equilibrium for 
selfish systems and it corresponds to the equilibrium induced by the 
{\em best-response} dynamics. 
The observation that selfish systems are described by their equilibrium
states (that is, by the Nash equilibria) has motivated the notions of 
Price of Anarchy \cite{KP} and Price of Stability \cite{PoS} and 
the analysis of efficiency of selfish systems based on such notions.

However, such analysis inherits some of the shortcomings 
of the concept of a Nash equilibrium. 
First of all, the best-response dynamics assumes that selfish agents
have complete knowledge of the current state of the system; that is, they know the
payoff associated with each of their possible choices and each of the strategies chosen by other 
agents.
Instead, in most cases, agents have only approximate knowledge of the system state or they are not able to compute their best choice.
Moreover, in presence of multiple equilibria, it is not clear which 
one of them will be reached by the system, as it 
may depend on the initial state:
Price of Anarchy considers the
worst-case equilibrium, whereas Price of Stability focuses on the best-case
equilibrium.
Finally, Nash equilibria are hard to compute \cite{nash1,nash2} and thus
for some systems it might take very long to reach a Nash equilibrium:
in this case using equilibrium states to describe the system 
performance is not well justified. 
Rather, one would like to analyze the performance of a system by using a 
dynamics (and its related equilibrium notion) 
that has the following three properties:
\begin{itemize}
\item The dynamics takes into account the fact that the system components might have 
a perturbed or noisy knowledge of the system;
\item For every system the equilibrium state exists and is unique;
\item The system reaches the equilibrium very quickly regardless of the starting state.
\end{itemize}

In this paper, we consider {\em \rbrdyn} in which 
the behavior of the agents is described by a parameter
$\beta\geq 0$. The case $\beta=0$ corresponds 
to agents picking their strategies completely at random
(that is, the agents have no knowledge of the system)
and the case $\beta=\infty$ corresponds to agents picking their strategies according to the 
best-response dynamics
(in which the agents have full and complete knowledge of the system). 
The intermediate values of $\beta$ correspond to agents
that are roughly guided by the best-response dynamics but can make 
a sub-optimal choice due, for example, to \emph{bounded rationality} of the agent or \emph{limited knowledge} about the system: this sub-optimal behavior occurs with some probability that depends on $\beta$
(and on the associated payoff).

We will study a specific \rbrdyn\ for which
the system evolves according to an ergodic Markov chain for all $\beta\geq 0$.
For these systems, 
it is natural to look at the stationary distribution 
(which is the equilibrium state of the Markov chain) 
and to analyze the expected social welfare (the sum of utility functions) of the system at that distribution.
We stress that the \rbrdyn\ well models agents that only have approximate or noisy
knowledge of the system and that
for ergodic Markov chains (such as the ones arising in our study)
the stationary distribution is known to exist and to be unique.
Moreover, to justify the use of the stationary distribution 
for analyzing the performance of the system, we will study how fast 
the Markov chain  converges to the stationary distribution.

\smallskip\noindent{\bf Related Works and Our Results.}
Several dynamics, besides the best-response dynamics,
and several notions of equilibrium, besides Nash equilibria,
have been considered to describe the evolution of a selfish system
and to analyze its performance. 
See, for example,~\cite{flMIT98,youngPUP98,sandholmMIT10}.

\smallskip\noindent{\em Equilibrium concepts based on the best-response.}
When the game does not possess a Pure Nash equilibrium, 
the best-response dynamics will eventually 
cycle over a set of states (in a Nash equilibrium the set is a singleton). 
These states are called {\em sink equilibria} \cite{gmvFOCS05}. 
Sink equilibria exist for all games and, 
in some contexts, they seem a better approximation of the real setting than 
{\em mixed} Nash equilibria.
Unfortunately, sink equilibria share two undesirable properties
with Nash equilibria: 
a game can have more that one sink equilibrium and sink equilibria 
seem hard to compute \cite{fpSODA08}.

Other notions of equilibrium state associated with best-response dynamics are the 
\textit{unit-recall equilibria} and \textit{component-wise unit-recall equilibria} 
(see \cite{fpSODA08}). 
However, we point out that the former does not always exist and 
that the latter imposes too strict limitations on the players. 

\smallskip\noindent{\em No-Regret Dynamics.}
Another broadly explored set of dynamics are the no-regret dynamics 
(see, for example, \cite{flMIT98}).
The regret of an user is the difference between the long-term average cost 
and the average cost of the best strategy in hindsight. 
In the no-regret dynamics the regret of every player after $t$ steps is 
$o(t)$ (sublinear with time). 
In \cite{fvGEB97,hmECO00} it is showed that the no-regret dynamics 
converges to the set of \emph{correlated equilibria}.
Note that the convergence is to the {\em set} of correlated equilibria 
and not to a specific correlated equilibrium.

\smallskip\noindent{\em Our work.}
In this paper we consider a specific \rbrdyn\ called the {\em logit} dynamics
(see \cite{blumeGEB93}) and we study its mixing time 
(that is, the time it takes to converge to the stationary distribution) and the stationary expected social welfare. 
Specifically, 
\begin{itemize}
	\item We start by analyzing the logit dynamics
	for a simple $3$-player linear congestion game 
	(the \ck\ game~\cite{ckSTOC05}) which
	exhibits the worst Price of Anarchy among linear congestion games.
	We show that the mixing time
	of the logit dynamics is upper bounded by a constant independent of $\beta$. 
	Moreover, 
	we show that the stationary expected social welfare is larger than the social welfare
	of the worst Nash equilibrium for all $\beta$;
	\item We then analyze the $2\times 2$ coordination games studied in \cite{blumeGEB93}.
	Here we show that, under some conditions, the stationary expected social welfare is larger than the social welfare of 
	the worst Nash equilibrium. 
	We give upper and lower bounds 
	on the mixing time exponential in $\beta$. 
	We also observe that the same bounds apply to anti-coordination games;
	\item Finally, 
		we apply our analysis to two simple $n$-player games: the OR game and XOR game. We give upper and lower bounds on the mixing time: we show that the mixing time of the OR game can be upper bounded by a function independent of $\beta$, while the mixing time of the XOR game increases exponentially in $\beta$. We also prove that for $\beta=\mathcal{O}(\log n)$ the mixing time is polynomial in $n$ for both games.
\end{itemize}
The {\em logit} dynamics %and the corresponding Markov chain
has been first studied by Blume~\cite{blumeGEB93} who showed 
that, for $2\times 2$ coordination games, 
the long-term behavior of the Markov chain is concentrated
in the risk dominant equilibrium (see~\cite{hsMIT88}) 
for sufficiently large $\beta$.
Ellison~\cite{ellisonECO93}  studied different \rbrdyn\ for coordination games 
assuming that interaction among players were described by a graph;
that is, the utility of a player is determined only by the strategies 
of the adjacent players. Specifically, 
Ellison~\cite{ellisonECO93} studied interaction 
modeled by rings and showed that some large
fraction of the players will eventually choose the risk dominant strategy.
Similar results were obtained by Peyton Young~\cite{youngTR00} for the logit dynamics
and for more general families of graphs.
Montanari and Saberi~\cite{msFOCS09} gave bounds on the hitting time (the expected time that the logit dynamics takes to reach a specific state) 
of the risk dominant equilibrium state
in terms of some graph theoretic properties of the underlying interaction network. 
Asadpour and Saberi~\cite{asWINE09} studied the hitting time for 
a broader class of congestion games.
We notice that none of \cite{blumeGEB93,ellisonECO93,youngTR00} gave any 
bound on the convergence time to the risk dominant equilibrium. 
Montanari and Saberi \cite{msFOCS09}
were the first to do so but their study focuses on the hitting time of a specific 
configuration and not on the convergence time to the stationary distribution.

From a technical point of view, our work follows the lead of 
\cite{blumeGEB93,ellisonECO93,youngTR00} and extends their technical 
findings by giving bounds on the mixing time of the Markov chain of the logit dynamics.
We stress that previous results only proved that, for sufficiently large $\beta$, {\em eventually} 
the system concentrates around certain states without further 
quantifying the rate of convergence nor the asymptotic behaviour of the system for small values of $\beta$.
Instead, we identify the stationary distribution of the logit dynamics as the \emph{global} equilibrium and we evaluate the social welfare at stationarity and the time it takes the system to reach it (the mixing time) as explicit functions of  $\beta$.

We choose to start our study from the class of coordination games 
considered in~\cite{blumeGEB93} and
two simple $n$-player games (the OR game and the XOR game). We give nearly tight upper and lower bounds 
on the mixing time. Despite their \emph{game-theoretic} simplicity, 
the analytical study of the mixing time of  the logit dynamics for the two $n$-player games is far from trivial.
%Pino: 
We notice that the results in~\cite{msFOCS09} cannot be used to derive upper
bounds on the mixing time.

From a more conceptual point of view, our work tries 
(similarly to \cite{gmvFOCS05,fpSODA08,bkmrSODA10})
to introduce a solution concept that well models the behavior of selfish agents, is
uniquely defined for any game, and is quickly reached from any starting state. We propose the 
stationary distribution induced by the logit dynamics 
as a possible solution concept and exemplify its use in the analysis of the performance
of some $2\times 2$ games (as the ones considered in \cite{blumeGEB93}), 
of games used to obtain tight bounds on the Price of Anarchy, and of two simple
multi-player games.

\paragraph{Organization of the paper.}
In Section~\ref{apx:markov} we summarize some Markov chain notions that we will use throughout the paper.
In Section~\ref{sec::model} 
we formally describe the logit dynamics for strategic games. 
We also describe the coupling we will repeatedly use in the proofs of the upper 
bounds on mixing times. 
In Sections~\ref{sec::ckgame}, \ref{sec::twoplayers}, \ref{sec::orgame} and \ref{sec:xorgame}
we study the stationary expected social welfare and the 
mixing time of the logit dynamics for \ck\ game, 
coordination games, the OR game, and the XOR game, respectively.
Finally, in Section~\ref{sec::conclusions} we present conclusions and some open problems.

\smallskip\noindent
{\em Notation.} 
We write $\overline{S}$ for the complementary set of a set $S$; 
we write $|S|$ for its size. 
We use bold symbols for vectors; when 
$\mathbf{x} = (x_1, \dots, x_n) \in \{0,1\}^n$ we write $|\mathbf{x}|$ for 
the number of $1$s in $\mathbf{x}$; i.e., 
$|\mathbf{x}|=|\{i \in [n] \,:\, x_i = 1 \}|$.
For two vectors $\mathbf{x}, \mathbf{y}$ let $\hamming(\x,\y) = |\{ i \in [n] \,:\, x_i \neq y_i \}|$ be their Hamming distance: we write $\mathbf{x} \sim \mathbf{y}$ if $\hamming(\x,\y) = 1$.
We use the standard game theoretic 
notation $(\mathbf{x}_{-i},y)$ to mean the vector obtained from $\mathbf{x}$ by replacing the $i$-th entry with $y$, i.e. $(\mathbf{x}_{-i},y) = (x_1, \dots, x_{i-1},y,x_{i+1}, \dots, x_n)$.

%% file: trunk/Markov.tex
We summarize the main tools we use to bound the mixing time of Markov chains
(for a complete description of such tools see, for example, Chapters~5.2, 7.2, 12.2 and 14.2 of~\cite{lpwAMS08}. We refer the reader to~\cite{lpwAMS08} also for notational conventions).

\smallskip \noindent
Consider a 
Markov chain $\mathcal{M}$ with {\em finite} state space $\Omega$ and 
transition matrix $P$. 
It is a classical result that for an \emph{irreducible} and \emph{aperiodic} Markov chain\footnote{Roughly speaking, a finite-state Markov chain is irreducible and aperiodic if there is a time $t$ such that, for all pairs of states $x,y$, the probability to be in $y$ after $t$ steps, starting from $x$, is positive.}
there exists an unique {\em stationary distribution} $\pi$ over $\Omega$;
that is, a distribution $\pi$ on $\Omega$ such that $\pi\cdot P=\pi$.

\noindent
The {\em total variation} distance 
$\|\mu - \nu \|_{\text{TV}}$
between two probability distributions 
$\mu$ and $\nu$ on $\Omega$ is defined as 
$$
\|\mu - \nu \|_{\text{TV}}=\max_{A\subset\Omega}
	|\mu(A)-\nu(A)|\,.
$$
An irreducible and aperiodic Markov chain $\mathcal{M}$
\emph{converges} to its stationary distribution $\pi$; 
specifically, there exists $1>\alpha>0$ such that
$$d(t)\leq\alpha^t,$$
where 
$$d(t)=
\max_{x\in\Omega}
\|P^t(x,\cdot) - \pi \|_{\text{TV}}$$
and 
$P^t(x,\cdot)$ is the distribution at time $t$ 
of the Markov chain starting at $x$. 
For $1/2>\varepsilon > 0$, 
the \emph{mixing time} is defined as
$$
t_{\text{mix}}(\varepsilon) = 
\min \{t\in\mathbb{N}\,:\, d(t)\leq\varepsilon\}.
$$
It is usual to set $\varepsilon = 1/4$ or $\varepsilon = 1/2e$. 
If not explicitly specified, when we write $\tm$ we mean $\tm(1/4)$.
Observe that 
$\tm(\epsilon)\leq\lceil\log_2 \epsilon^{-1}\rceil\tm$.

\paragraph{Coupling.}
A {\em coupling} of two probability distributions $\mu$ and $\nu$  on 
$\Omega$ is a pair of random variables $(X,Y)$ defined on 
$\Omega\times\Omega$ such that the marginal distribution of 
$X$ is $\mu$ and the marginal distribution of $Y$ is $\nu$.
A {\em coupling of a Markov chain} $\mathcal{M}$ 
with transition matrix $P$ is a process 
$(X_t,Y_t)_{t=0}^\infty$ with the property that 
both $X_t$ and $Y_t$ are Markov chains with transition matrix $P$. When the two coupled chains start at $(X_0,Y_0) = (x,y)$, we write $\Prob{x,y}{\cdot}$ and $\Expec{x,y}{\cdot}$ for the probability and the expectation on the space where the two chains are both defined.

\noindent
We denote  by $\tc$ the first time the two chains meet; that is, 
$$
\tc=\min\{t: X_t=Y_t\}\,.
$$
We will consider only couplings of Markov chains with the property that 
for $s\geq\tc$, it holds $X_s=Y_s$.
The following theorem can be used to give an upper bound on $\tm$
(see, for example, Corollary 5.3 in \cite{lpwAMS08}).
\begin{theorem}[Coupling]
\label{thm:coupling}
Let $\mathcal{M}$ be a Markov chain with state space $\Omega$ and 
transition matrix $P$. For each pair of states $x,y\in\Omega$ consider a coupling $(X_t,Y_t)$ of $\mathcal{M}$ with starting states $X_0=x$ and $Y_0=y$.
Then
$$
d(t)\leq\max_{x,y\in\Omega} \Prob{x,y}{\tc>t}\,.
$$
\end{theorem}
Sometimes it is difficult to specify a coupling and to 
analyze the coupling time $\tc$ 
for each pair of starting states $x$ and $y$.
The \emph{Path Coupling} theorem says that it is sufficient to define a coupling 
only for pairs of Markov chains starting from \emph{adjacent} states and an upper bound on the mixing
time can be obtained if each of these couplings contracts their distance 
on average.
More precisely, consider a  Markov chain $\mathcal{M}$ 
with state space $\Omega$ and transition matrix $P$;
let $G=(\Omega,E)$ be a connected graph and let $w:E\rightarrow\mathbb{R}$
be  a function assigning weights to the edges such that 
$w(e)\geq 1$ for every edge $e\in E$; for $x,y\in\Omega$, 
we denote by $\rho(x,y)$ the weight of the shortest path 
in $G$ between $x$ and $y$.  The following theorem holds.

\begin{theorem}[Path Coupling~\cite{BubleyDyer}]
\label{theorem:pathcoupling}
Suppose that for every edge $\{x,y\}\in E$ a coupling
$(X_t,Y_t)$ of 
$\mathcal{M}$ with $X_0=x$ and $Y_0=y$ exists such that 
$\Expec{x,y}{\rho(X_1,Y_1)} \leqslant e^{-\alpha}\cdot w(\{x,y\})$ 
for some $\alpha > 0$. 
Then 
%the mixing time $t_{\text{mix}}(\varepsilon)$ of $\mathcal{M}$ is
$$
t_{\text{mix}}(\varepsilon) \leqslant \frac{\log(\text{diam}(G)) + \log(1/\varepsilon)}{\alpha}
$$
where $\text{diam}(G)$ is the (weighted) diameter of $G$.
\end{theorem}

\paragraph{Spectral techniques.}
A Markov chain $\mathcal{M}$ with state space $\Omega$ and 
transition matrix $P$ is said \emph{reversible} if for all $x,y\in\Omega$, 
it holds that
$$\pi(x)\cdot P(x,y)=\pi(y)\cdot P(y,x).$$
The eigenvalues of the transition matrix $P$ 
of a reversible Markov 
chain $\mathcal{M}$ can be used to obtain upper and lower bounds on the 
mixing time.
Observe that all the eigenvalues of any transition matrix $P$ 
have absolute value at most $1$, $\lambda = 1$ is an eigenvalue, 
and for irreducible and aperiodic chains, $-1$ is not an eigenvalue. 
The {\em relaxation time} 
$t_{\text{rel}}$ of a Markov chain $\mathcal{M}$ is defined as
$$
t_{\text{rel}}={\frac{1}{1-\lambda^\star}}
$$
where $\lambda^\star$ is the largest absolute value among eigenvalues 
other than $1$,
$$
\lambda^\star = \max\{ |\lambda| \;:\; \lambda \text{ is an eigenvalue of } P, \, \lambda \neq 1 \} \,.
$$
Observe that, for $\mathcal{M}$ reversible, irreducible and aperiodic,
$0\leq\lambda^\star<1$ and thus
$t_{\text{rel}}$ is positive and finite. We have the following theorem
(see, for example, Theorems 12.3 and 12.4 in \cite{lpwAMS08}).
\begin{theorem}[Relaxation time]\label{theorem:relaxation}
Let $P$ be the transition matrix of a reversible, irreducible, and aperiodic Markov chain with state space 
$\Omega$ and stationary distribution $\pi$. Then 
$$
(t_{\text{rel}}-1)\log\left({\frac{1}{2\epsilon}}\right)
\leq \tm(\epsilon)\leq
\log\left({\frac{1}{\epsilon\pi_{\text{min}}}}\right) t_{\text{rel}}
$$
where 
$\pi_{\text{min}}=\min_{x\in\Omega} \pi(x)$.
\end{theorem}

\paragraph{Lower bound.} We will use the following theorem to derive our
lower bounds (see, for example, Theorem~7.3 in \cite{lpwAMS08}).
\begin{theorem}[Bottleneck ratio]\label{theorem:bottleneck}
Let $\mathcal{M} = \{ X_t \,:\, t \in \mathbb{N} \}$ be an irreducible and aperiodic Markov 
chain with finite state space $\Omega$, 
transition matrix $P$, and stationary distribution $\pi$. 
Let $S \subseteq \Omega$ be any set with $\pi(S) \leqslant 1/2$. 
Then the mixing time is
$$
t_{\text{mix}}(\varepsilon) \geqslant \frac{1-2\epsilon}{2 \Phi(S)}
$$
where
$$
\Phi(S) = \frac{Q(S,\overline{S})}{\pi(S)} \quad \mbox{ and } \quad Q(S, \overline{S}) = \sum_{x \in S, \, y \in \overline{S}} \pi(x) P(x,y).
$$
\end{theorem}

%% file: trunk/model.tex
A \emph{strategic game} is a triple $([n], \mathcal{S}, \mathcal{U})$, where $[n] = \{1,\dots,n\}$ is a finite set of \emph{players}, $\mathcal{S}=\{S_1,\dots,S_n\}$ is a family of non-empty finite sets ($S_i$ is the set of \emph{strategies} for player $i$), and $\mathcal{U} = \{u_1,\dots,u_n\}$ is a family of \emph{utility functions} (or \emph{payoffs}), where $u_i \,: \, S_1 \times \cdots \times S_n \rightarrow \mathbb{R}$ is the utility function of player $i$.

\smallskip \noindent
Consider the following \emph{noisy} best-response dynamics, introduced in~\cite{blumeGEB93} and known as \emph{logit dynamics}: at every time step
\begin{enumerate}
\item Select one player $i \in [n]$ uniformly at random;
\item Update the strategy of player $i$ according to the following probability distribution over the set $S_i$ of her strategies. For every $y \in S_i$
\begin{equation}\label{eq:updateprob}
\sigma_i(y \,|\, \mathbf{x}) = \frac{1}{T_i(\mathbf{x})} \, e^{\beta u_i(\mathbf{x}_{-i}, y)}
\end{equation}
where $\mathbf{x} \in S_1 \times \cdots \times S_n$ is the strategy profile played at the current time step, 
$T_i(\mathbf{x}) \linebreak = \sum_{z \in S_i} e^{\beta u_i(\mathbf{x}_{-i}, z)}$ is the normalizing factor, and $\beta \geqslant 0$.
\end{enumerate}
Parameter $\beta$ is called \emph{inverse noise} of the system, indeed from~(\ref{eq:updateprob}) it is easy to see that, for $\beta = 0$ player $i$ selects her strategy uniformly at random, for $\beta > 0$ the probability is biased toward strategies promising higher payoffs, and for $\beta \rightarrow \infty$ player $i$ chooses her best response strategy (if more than one best response is available, she chooses uniformly at random one of them). Moreover observe that probability $\sigma_i(y \,|\, \mathbf{x})$ does not depend on the strategy $x_i$ currently adopted by player $i$.

The above dynamics defines a Markov chain with the set of  strategy profiles as state space, and where the transition probability from profile $\mathbf{x} = (x_1, \dots, x_n)$ to profile $\mathbf{y} = (y_1, \dots, y_n)$ is zero if the $\hamming(\x,\y)\geqslant 2$ and it is $\frac{1}{n} \sigma_i(y_i \,|\, \mathbf{x})$ if the two profiles differ exactly at player $i$. More formally, we can define the logit dynamics as follows.

\begin{definition}[Logit dynamics~\cite{blumeGEB93}]
Let $\mathcal{G} = ([n], \mathcal{S}, \mathcal{U})$ be a strategic game and let $\beta \geqslant 0$. The \emph{logit dynamics} for $\mathcal{G}$ is the Markov chain $\mathcal{M}_\beta = \{ X_t \,:\, t \in \mathbb{N} \}$ with state space $\Omega = S_1 \times \cdots \times S_n$ and transition matrix
\begin{equation}\label{eq:transmatrix}
P(\mathbf{x}, \mathbf{y}) = \frac{1}{n} \cdot 
\left\{
\begin{array}{cl}
\sigma_i(y_i \,|\, \mathbf{x}), & \quad \mbox{ if } \mathbf{y}_{-i} = \mathbf{x}_{-i} \mbox{ and } y_i \neq x_i; \\[2mm]
\sum_{i=1}^n \sigma_i(y_i \,|\, \mathbf{x}), & \quad \mbox{ if } \mathbf{y} = \mathbf{x}; \\[2mm]
0, & \quad \mbox{ otherwise;}
\end{array}
\right.
\end{equation}
where $\sigma_i(y_i \,|\,\mathbf{x})$ is defined in (\ref{eq:updateprob}).
\end{definition}

\paragraph{Properties.} Logit dynamics enjoys some interesting properties: 

\smallskip\noindent
\textit{Ergodicity.} It is easy to see that the logit dynamics is irreducible and aperiodic. Indeed, let $\mathbf{x} = (x_1, \dots, x_n)$ and $\mathbf{y} = (y_1, \dots, y_n)$ be two profiles and let $(\mathbf{z}^0, \dots,  \mathbf{z}^n)$ be a \emph{path} of profiles where $\mathbf{z}^0 = \mathbf{x}, \, \mathbf{z}^n = \mathbf{y}$ and $\mathbf{z}^i = (y_1, \dots, y_i, x_{i+1}, \dots x_n)$ for $i = 1, \dots, n-1$. The probability that the chain starting at $\mathbf{x}$ is in $\mathbf{y}$ after $n$ steps is
$$
P^n(\mathbf{x}, \mathbf{y}) = P^n(\mathbf{z}^0, \mathbf{z}^n) \geqslant P^{n-1}(\mathbf{z}^0, \mathbf{z}^{n-1})P(\mathbf{z}^{n-1},\mathbf{z}^n) 
$$
and recursively
$$
P^n(\mathbf{x}, \mathbf{y}) \geqslant \prod_{i=1}^n P(\mathbf{z}^{i-1},\mathbf{z}^i) > 0
$$
where the last inequality follows from (\ref{eq:transmatrix}) because, for all $i=1,\dots, n$, the Hamming distance between $\mathbf{z}^{i-1}$ and $\mathbf{z}^i$ is at most $1$. Hence there is a unique stationary distribution $\pi$ and, for every starting profile $\x$, the distribution of the chain $P^t(\x, \cdot)$ converges to $\pi$ in total variation as $t$ tends to infinity.

\smallskip\noindent
\textit{Invariance under utility translation.} Let $\mathcal{G}= ([n],\mathcal{S},\mathcal{U})$ be a game. If we change the utility functions by adding a constant $c_i$ to all the utilities of player $i$, i.e. if we define a new family $\tilde{\mathcal{U}} = \{\tilde{u}_i \,:\, i \in [n]\}$ of utility functions as follows
$$
\tilde{u}_i(\mathbf{x}) := u_i(\mathbf{x})  + c_i \qquad \mbox{ for all } \mathbf{x}
$$
we get a new game $\tilde{\mathcal{G}}= ([n],\mathcal{S},\tilde{\mathcal{U}})$ but the same logit dynamics. Indeed, according to (\ref{eq:updateprob}), the probability player $i$ chooses strategy $y$ when the game is at profile $\mathbf{x}$ is
$$
\tilde{\sigma}_i(y\,|\,\x) = \frac{e^{\beta \tilde{u}_i(\mathbf{x}_{-i}, y)}}{\sum_{z \in S_i} e^{\beta \tilde{u}_i(\mathbf{x}_{-i}, z)}} = \frac{1}{\sum_{z \in S_i} e^{\beta \left[\tilde{u}_i(\mathbf{x}_{-i}, z) - \tilde{u}_i(\mathbf{x}_{-i}, y) \right]}} = \frac{1}{\sum_{z \in S_i} e^{\beta \left[u_i(\mathbf{x}_{-i}, z) - u_i(\mathbf{x}_{-i}, y) \right]}} = \sigma_i(y\,|\,\x)\,.
$$

\smallskip\noindent
\textit{Noise changes under utility rescaling.} While translations of utilities do not affect logit dynamics, a \emph{rescaling} of the utility functions for the same constant $\alpha>0$ changes the inverse noise from $\beta$ to $\alpha \cdot \beta$. Indeed, if for every player $i$ and every profile $\x$ we set
$$
\tilde{u}_i(\mathbf{x}) := \alpha \cdot u_i(\mathbf{x})\,,
$$
from (\ref{eq:updateprob}) we have
$$
\tilde{\sigma}_i(y\,|\,\x) = \frac{e^{\beta \tilde{u}_i(\mathbf{x}_{-i}, y)}}{\sum_{z \in S_i} e^{\beta \tilde{u}_i(\mathbf{x}_{-i}, z)}} = \frac{e^{\alpha\beta u_i(\mathbf{x}_{-i}, y)}}{\sum_{z \in S_i} e^{\alpha\beta u_i(\mathbf{x}_{-i}, z)}}\,.
$$
Notice that, unlike translations constants, we here must have the same rescaling constant $\alpha$ for all utility functions. 

\smallskip
\paragraph{Potential Games.}
A game $\mathcal{G} = ([n],\mathcal{S},\mathcal{U})$ is said a (exact) \emph{potential} game if a function $\Phi\,:\, S_1 \times \cdots \times S_n \rightarrow \mathbb{R}$ exists such that, for every player $i$ and for every pair of profiles $\x$ and $\y$ that differ only at position $i$, it holds that $u_i(\x) - u_i(\y) = \Phi(\x) - \Phi(\y)$.
It is easy to see that, if $\mathcal{G} = ([n], \mathcal{S}, \mathcal{U})$ is a potential game with potential function $\Phi$, then the Markov chain given by (\ref{eq:transmatrix}) is reversible and its stationary distribution is the Gibbs measure
\begin{equation}\label{eq:Gibbs}
\pi(\mathbf{x}) = \frac{1}{Z} e^{\beta \Phi(\mathbf{x})}
\end{equation}
where 
$Z = \sum_{\mathbf{y} \in S_1 \times \cdots \times S_n} e^{\beta \Phi(\mathbf{y})}$ 
is the normalizing constant. Except for the Matching Pennies example in Subsection~\ref{ssec::examples}, all the games we analyze in this paper are potential games.

\smallskip\noindent
\textit{Logit dynamics vs Glauber dynamics.}
When $\mathcal{G}$ is a potential game, the logit dynamics is equivalent to the well-studied \emph{Glauber dynamics}. For state space $\Omega = S_1 \times \cdots \times S_n$ and probability distribution $\mu$ over $\Omega$, the Glauber dynamics for $\mu$ proceeds as follows: From profile $\x \in \Omega$, pick a player $i \in [n]$ uniformly at random and update her strategy at $y \in S_i$ with probability $\mu$ conditioned on the other players being at $\x_{-i}$, i.e.
$$
\mu(y \,|\,\x_{-i}) = \frac{\mu(\x_{-i},y)}{\sum_{z \in S_i} \mu(\x_{-i},z)}\,.
$$
It is easy to see that the Markov chain defined by the Glauber dynamics is irreducible, aperiodic, and reversible with stationary distribution $\mu$. When $\mathcal{G} = ([n],\mathcal{S},\mathcal{U})$ is a potential game with potential function $\Phi$, the logit dynamics defines the same Markov chain as the Glauber dynamics for the Gibbs distribution $\pi$ in (\ref{eq:Gibbs}). Indeed, in that case we have
\begin{eqnarray*}
\sigma_i(y \,|\, \mathbf{x}) & = & \frac{e^{\beta u_i(\x_{-i},y)}}{\sum_{z \in S_i}e^{\beta u_i(\x_{-i},z)}}
= \frac{1}{\sum_{z \in S_i} e^{\beta \left( u_i(\x_{-i},z) - u_i(\x_{-i},y) \right)}} \\
& = & \frac{1}{\sum_{z \in S_i} e^{\beta \left( \Phi(\x_{-i},z) - \Phi(\x_{-i},y) \right)}}
=  \frac{e^{\beta \Phi(\x_{-i},y)}}{\sum_{z \in S_i}e^{\beta \Phi(\x_{-i},z)}}
= \frac{\pi(\x_{-i},y)}{\sum_{z \in S_i} \pi(\x_{-i},z)}\,.
\end{eqnarray*}
Hence, logit dynamics for potential games and Glauber dynamics for Gibbs distributions are two ways of looking at the same Markov chains: in the former case the dynamics is derived from the potential function, in the latter case from the stationary distribution. However, observe that, if $\mathcal{G}$ is not a potential game and $\pi$ is the stationary distribution of the logit dynamics for $\mathcal{G}$, in general the Glauber dynamics for $\pi$ is different from the logit dynamics (see, for example, the Matching Pennies case in Subsection~\ref{ssec::examples}).  

Due to the analogies between logit and Glauber dynamics, we will sometimes adopt the terminology used by physicists to indicate the quantities involved; in particular we will call parameter $\beta$ the inverse noise or \emph{inverse temperature} and we will call \emph{partition function} the normalizing constant $Z$ of the Gibbs distribution (\ref{eq:Gibbs}).

\paragraph{Stationary expected social welfare and mixing time.}
Let $W:S_1 \times \cdots \times S_n \longrightarrow \mathbb{R}$ be a \emph{social welfare function} (in this paper we assume that $W$ is simply the sum of all the utility functions $W(\mathbf{x}) = \sum_{i=1}^n u_i(\mathbf{x})$, but clearly any other function of interest can be analysed). We study the \emph{stationary expected social welfare}, i.e. the expectation of $W$ when the strategy profiles are random according 
to the stationary distribution $\pi$ of the Markov chain,
$$
\Expec{\pi}{W} = \sum_{\mathbf{x} \in S_1 \times \cdots \times S_n} W(\mathbf{x}) \pi(\mathbf{x})
$$

\noindent
Since the Markov chain defined in (\ref{eq:transmatrix}) is irreducible and aperiodic, from every initial profile $\textbf{x}$ the distribution $P^t(\mathbf{x}, \cdot)$ of chain $X_t$ starting at $\textbf{x}$ will eventually converge to $\pi$ as $t$ tends to infinity. We will be interested in bounding how long it takes to get close to the stationary distribution, that is the \emph{mixing time} of the Markov chain.

\ignore{
More formally, we define
$$
t_{\text{mix}}(\varepsilon) = 
\min_{t \in \mathbb{N}} 
	\max_{\mathbf{x} \in \Omega} 
	\left\{ \| P^t(\mathbf{x}, \cdot) - \pi  \|_{\text{TV}} \leqslant \varepsilon \right\}
$$
where $\|P^t(\mathbf{x},\cdot)-\pi\|_{\text{TV}} = 
	\frac{1}{2} \sum_{\mathbf{y} \in \Omega} |P^t(\mathbf{x}, \mathbf{y})-
	\pi(\mathbf{y})|$ is the \emph{total variation distance},
and we set 
$t_{\text{mix}}=t_{\text{mix}}(1/4)$.
}

\noindent
In the next subsection we illustrate the goals of our work with two simple examples. 

\subsection{Two simple examples: Matching Pennies and a \emph{Stairs} game}\label{ssec::examples}
\paragraph{Matching Pennies.} Consider the classical \emph{Matching Pennies} game. We write the utility functions in the standard bimatrix form. 

\begin{equation}\label{eq:mpdef}
 \begin{array}{ | c || c | c | }
  \hline
   & H & T \\
  \hline \hline
  H & +1,\, -1 & -1,\, +1 \\
  \hline
  T & -1,\, +1 & +1,\, -1 \\
  \hline
\end{array}
\end{equation}

\noindent
According to (\ref{eq:updateprob}), the update probabilities for the logit dynamics are, for every $x \in \{ H, \, T\}$
\begin{eqnarray*}
\sigma_1(H\,|\,(x,H)) = \sigma_1(T\,|\,(x,T)) = & \frac{1}{1 + e^{-2 \beta}} & = \sigma_2(T\,|\,(H,x)) = \sigma_2(H \,|\, (T,x))\,, \\[2mm]
\sigma_1(T \,|\, (x,H)) = \sigma_1(H \,|\, (x,T)) = & \frac{1}{1 + e^{2 \beta}} & = \sigma_2(H \,|\,(H,x)) = \sigma_2(T \,|\, (T,x))\,.
\end{eqnarray*}
Hence the transition matrix (see (\ref{eq:transmatrix})) is
$$
P =
\left(
\begin{array}{c|cccc}
& HH & HT & TH & TT \\
\hline
HH & 1/2 & b/2 & (1-b)/2 & 0 \\[1mm]
HT & (1-b)/2 & 1/2 & 0 & b/2 \\[1mm]
TH & b/2 & 0 & 1/2 & (1-b)/2 \\[1mm]
TT & 0 & (1-b)/2 & b/2 & 1/2
\end{array}
\right)
$$
where, for readability sake, we named $b = \frac{1}{1+e^{-2\beta}}$.

Since every column of the matrix adds up to $1$, the uniform distribution $\pi$ over the set of strategy profiles is the stationary distribution for the logit dynamics. The stationary expected social welfare is thus $0$ for every inverse noise $\beta$.

As for the mixing time, it is easy to see that it is upper bounded by a constant independent of $\beta$. Indeed, a direct calculation shows that, for every $\mathbf{x} \in \{HH, \, HT, \, TH, \, TT \}$ and for every $\beta \geqslant 0$ it holds that
$$
\| P^3(\mathbf{x},\cdot) - \pi \|_{\text{\textsc{tv}}} \leqslant \frac{7}{16} < \frac{1}{2}\,.
$$

\paragraph{A \emph{stairs} game.}
One of the main techniques used to give upper bounds on the mixing time of Markov chains is the coupling technique (see Theorem~\ref{thm:coupling}). In the following example we use it to upper bound the mixing time of the logit dynamics for a simple game.

Let $\mathcal{G}$ be a potential game where every player has two strategies, say \emph{upstairs} (or $1$)  and \emph{downstairs} (or $0$), and the potential of a profile $\mathbf{x} \in \{0,1\}^n$ is the number of players that are upstairs, i.e. $\Phi(\mathbf{x}) = |\mathbf{x}|$.

Notice that the logit dynamics (and thus the stationary distribution and the mixing time) is completely defined by the potential function, while if we wanted to evaluate the stationary expected social welfare we would need to specify the utility functions.

\smallskip\noindent
The partition function is
$$
Z(\beta) = \sum_{\mathbf{x} \in \{0,1\}^n} e^{\beta |\mathbf{x}|} = \sum_{k=0}^n \binom{n}{k} e^{\beta k} = \left(1 + e^{\beta}\right)^n\,.
$$
So the stationary distribution is
$$
\pi(\mathbf{x}) = \frac{e^{\beta|\mathbf{x}|}}{\left(1 + e^{\beta}\right)^n}\,.
$$
As for the mixing time, we can use the coupling technique as follows: observe that the probability of playing strategy $1$ (or equivalently strategy $0$), for the player selected for the update, is independent of the current strategies of the other players. Indeed, according to~(\ref{eq:updateprob}), for every $\x$ it holds that
\begin{eqnarray*}
\sigma_i(1 \,|\, \mathbf{x})
& = & \frac{e^{\beta u_i(\mathbf{x}_{-i}, 1)}}{e^{\beta u_i(\mathbf{x}_{-i}, 1)} + e^{\beta u_i(\mathbf{x}_{-i}, 0)}}
= \frac{1}{1 + e^{\beta \left(u_i(\mathbf{x}_{-i}, 0) - u_i(\mathbf{x}_{-i}, 1) \right)}} \\
& = & \frac{1}{1 + e^{\beta \left(\Phi(\mathbf{x}_{-i}, 0) - \Phi(\mathbf{x}_{-i}, 1) \right)}}
= \frac{1}{1 + e^{\beta \left( |\mathbf{x}_{-i}| - (|\mathbf{x}_{-i}| + 1) \right)}}
= \frac{1}{1 + e^{-\beta}}\,.
\end{eqnarray*}
We can define a coupling of two Markov chains starting at two different profiles as follows: choose $i \in [n]$ uniformly at random and perform the same update at player $i$ in both chains\footnote{This is the same coupling used in the analysis of the lazy random walk on the hypercube (e.g. see~Section 5.3.3 in~\cite{lpwAMS08}), the only difference being that the probability of choosing $0$ or $1$ is not $1/2,1/2$ but $1/(1+e^{\beta}), \, 1/(1+e^{-\beta})$}. When every player has been chosen at least once the two chains have coalesced. From the coupon collector's argument, it takes $\mathcal{O}(n \log n)$ to have that, with probability at least $3/4$, all players have been chosen at least once. By applying Theorem~\ref{thm:coupling} we have that the mixing time is $\mathcal{O}(n \log n)$.

\smallskip
In the above examples, it turned out that the mixing time of the logit dynamics can be upper bounded by functions that do not depend on the inverse noise $\beta$. As we shall see in the next sections, this is not always the case. Moreover, the analysis of the mixing time is usually far from trivial.

%% file: trunk/coupling.tex
\subsection{Description of the Coupling}
\label{sec:coupling}
Throughout the paper we will use the coupling and path-coupling techniques 
(see Theorem~\ref{thm:coupling} and Theorem~\ref{theorem:pathcoupling}) to give upper bounds on mixing times. Since we will use the same coupling idea in several proofs, we describe it here and we will refer to this description when we will need it.

Consider an $n$-player $2$-strategy game $\mathcal{G}$ and let us rename $0$ and $1$ the strategies of every player. For every pair of strategy profiles $\x = (x_1, \dots, x_n), \y = (y_1, \dots, y_n) \in \{0,1\}^n$ we define a coupling $(X_1,Y_1)$ of two copies of the Markov chain with transition matrix $P$ defined in (\ref{eq:transmatrix}) for which $X_0 = \x$ and $Y_0 = \y$.

The coupling proceeds as follows: first, pick a player $i$ uniformly at random; then, update the strategies $x_i$ and $y_i$ of player $i$ in the two chains, by setting
$$(x_i,y_i)=\begin{cases}
	(0,0), & \text{with probability }
		\min\{\sigma_i(0 \mid \x),\sigma_i(0 \mid \y)\}\,;\cr
	(1,1), & \text{with probability }
		\min\{\sigma_i(1 \mid \x),\sigma_i(1 \mid \y)\}\,;\cr
	(0,1), & \text{with probability }
		\sigma_i(0 \mid \x)-\min\{\sigma_i(0 \mid \x),\sigma_i(0 \mid \y)\}\,;\cr
	(1,0), & \text{with probability }
		\sigma_i(1 \mid \x)-\min\{\sigma_i(1 \mid \x),\sigma_i(1 \mid \y)\}\,.
	\end{cases}$$
Three easy observations are in order: if $\sigma_i(0 \mid \x)=\sigma_i(0 \mid \y)$ and player $i$ is chosen, then, after the update, we have $x_i=y_i$; for every player $i$, at most one of the updates $(x_i,y_i)=(0,1)$ and $(x_i,y_i)=(1,0)$ has positive probability; if $i$ is chosen for update, then the marginal distributions of $x_i$ and $y_i$ agree with $\sigma_i(\cdot \mid \x)$ and $\sigma_i(\cdot \mid \y)$ respectively, indeed, for $b \in \{0,1\}$, the probability that $x_i = b$ is
$$
\min\left\{
\sigma_i(b \mid \x),\sigma_i(b \mid \y)
\right\} + 
\sigma_i(b \mid \x)-\min\{\sigma_i(b \mid \x),\sigma_i(b \mid \y)\}=
\sigma_i(b \mid \x)\,,
$$
and the probability that $y_i = b$ is
\begin{eqnarray*}
&& \min\{\sigma_i(b \mid \x),\sigma_i(b \mid \y)\} + \sigma_i(1 - b \mid \x) - \min\{\sigma_i(1 - b \mid \x),\sigma_i(1 - b \mid \y)\} =\\
 && = \min\{\sigma_i(b \mid \x),\sigma_i(b \mid \y)\ + (1 - \sigma_i(b \mid \x)) - (1 - \max\{\sigma_i(b \mid \x),\sigma_i(b \mid \y)\}) = \sigma_i(b \mid \y)\,.
\end{eqnarray*}

We define $G = (\Omega, E)$ as the \emph{Hamming graph} of the game, where $\Omega = \{0,1\}^n$ is the set of strategy profiles, and two profiles $\mathbf{x} = (x_1, \dots, x_n), \mathbf{y} = (y_1, \dots, y_n) \in \Omega$ are adjacent if they differ only for the strategy of one player, i.e.
\begin{equation}\label{eq:Hamming}
\{ \mathbf{x},\mathbf{y} \} \in E \; \Longleftrightarrow \;
% \left| \left\{ j \in [n] \;:\; x_j \neq y_j \right\} \right| = 1\,.
\x \sim \y\,.
\end{equation}
For the path coupling technique, the coupling described above is applied only to pairs of adjacent starting profiles.

%% file: trunk/ckgame.tex
In this section we analyze the \ck\ game, a simple $3$-player linear congestion game introduced in \cite{ckSTOC05}. This game is interesting because it highlights the weakness of the Price of Anarchy notion for the logit dynamics. Indeed, the \ck\ game exhibits the worst Price of Anarchy with respect to the average social welfare among all linear congestion games with 3 or more players. But, as we shall see soon, the stationary expected social welfare of the logit dynamics is 
always larger than the social welfare of the worst Nash equilibrium and, 
for large enough $\beta$, players spend most of the time in the best 
Nash equilibrium. Moreover, we will show that the mixing time of 
the logit dynamics can be bounded independently from $\beta$: that is,
the stationary distribution guarantees a good social welfare and it
is quickly reached by the system.

Let us now describe the \ck\ game.
We have $3$ players and $6$ facilities divided into two sets: 
$G=\{g_1, g_2, g_3\}$ and $H=\{h_1, h_2, h_3\}$. 
Player $i\in\{0,1,2\}$ has two strategies: 
Strategy ``$0$'' consists in selecting facilities $(g_i, h_i)$;
Strategy ``1'' consists in selecting facilities 
$(g_{i+1}, h_{i-1}, h_{i+1})$ 
(index arithmetic is modulo $3$).
The cost of a facility is the number of players choosing such facility, and the welfare of a player is minus the sum of the costs of the facilities she selected.
It easy to see that this game has two pure Nash equilibria: 
the solution where every player plays strategy $0$ 
(each player pays $2$, which is optimal), 
and the solution where every player plays strategy $1$ (each player pays $5$). 
The game is a congestion game, and thus, by~\cite{rosenthal73}, it is also a potential game and its potential function is:
$$
\Phi(\mathbf{x}) = \sum_{j \in G \cup H} \sum_{i = 1}^{L_\mathbf{x}(j)} i\,,
$$
where $L_\mathbf{x}(j)$ is the number of players using facility $j$ in configuration $\mathbf{x}$.

\paragraph{Stationary expected social welfare.}
It is easy to see that the update probabilities given by the logit dynamics for this game (see Equation~(\ref{eq:updateprob})) only depend on the number of players playing strategy 1 and not on which player is actually playing that strategy. In particular, we have that, from a profile $\x$, the player $i$, if selected for update, plays strategy 0 with the following probabilities:
\begin{equation}
 \label{eq:update_ck}
 \sigma_i( 0 \mid |\mathbf{x}_{-i}| = 0 ) = \frac{1}{1 + e^{-4\beta}}\,, \qquad
 \sigma_i( 0 \mid |\mathbf{x}_{-i}| = 1 ) = \frac{1}{1 + e^{-2\beta}}\,, \qquad
 \sigma_i( 0 \mid |\mathbf{x}_{-i}| = 2 ) = \frac{1}{2}\,,
\end{equation}
and strategy 1 with the remaining probabilities.

Next theorem evaluates the stationary expected social welfare for this game.
\begin{theorem}[Expected social welfare]
 The stationary expected social welfare $\Expec{\pi}{W}$ of the logit dynamics for the \ck\ game is
 $$
  \Expec{\pi}{W} = - \frac{6 + 39 e^{-4\beta} + 63 e^{-6\beta}}{1 + 3 e^{-4\beta} + 4 e^{-6\beta}}\,.
 $$
\end{theorem}
\proof
We notice that two profiles with the same number of players playing strategy $1$ have both the same potential (and, by Equation~(\ref{eq:Gibbs}), the same stationary distribution) and the same social welfare. Thus,  $\pi(\x) = \pi[k]$ and $W(\x) = W[k]$ for a profile $\x$ such that $|\x| = k$, where
$$
 \pi[0] = \frac{e^{-6\beta}}{Z(\beta)}\,, \qquad \pi[1] = \frac{e^{-10\beta}}{Z(\beta)}\,, \qquad \pi[2] = \pi[3] = \frac{e^{-12\beta}}{Z(\beta)}\,,
$$
where $Z(\beta) = e^{-6\beta} + 3e^{-10\beta} + 4e^{-12\beta}$, and
$$
 W[0] = -6\,, \qquad W[1] = -13\,, \qquad W[2] = -16\,, \qquad W[3] = -15\,.
$$
Hence, the stationary expected social welfare is
\begin{equation*}
\Expec{\pi}{W} = - \frac{6 \cdot e^{-6\beta} + 3 \cdot 13 \cdot e^{-10\beta} + (3 \cdot 16 + 15) \cdot e^{-12\beta}}{e^{-6\beta} + 3e^{-10\beta} + 4e^{-12\beta}} = - \frac{6 + 39 e^{-4\beta} + 63 e^{-6\beta}}{1 + 3 e^{-4\beta} + 4 e^{-6\beta}}\,.
\end{equation*}
\qed

Notice that for $\beta = 0$ we have $\Expec{\pi}{W} = - 27/2$, which is better than the social welfare of the worst Nash equilibrium. This means that, even if each player selects her strategy at random, the logit dynamics drives the system to a random profile whose expectation to be better than the worst Nash equilibrium. We also observe that $\Expec{\pi}{W}$ increases with $\beta$ and thus the long-term behavior of the logit dynamics gives a better social welfare than the worst Nash equilibrium for any $\beta \geqslant 0$. Moreover, the stationary expected social welfare approaches the optimal social welfare as $\beta$ tends to $\infty$.

\paragraph{Mixing time.}
Now we study the mixing time of the logit dynamics for the \ck\ game and we show that it is bounded by a constant for any $\beta\geqslant 0$. The proof will use Coupling Theorem (see Theorem~\ref{thm:coupling}).

\begin{theorem}[Mixing time]\label{theorem:ckgamesmain}
There exists a constant $\tau$ such that 
the mixing time $t_{\text{mix}}$ of the 
logit dynamics of the \ck\ game is upper bounded by $\tau$ 
for every $\beta\geq 0$.
\end{theorem}
\proof
First, we notice that the update probabilities given in Equation~(\ref{eq:update_ck}) imply that
\begin{equation}
 \label{eq:up_prop}
 \forall\,i, \forall\,\x, \forall\,\beta, \qquad \sigma_i(0 \,|\, \mathbf{x}) \geqslant 1/2\,.
\end{equation}
Let $X_t$ and $Y_t$ be two copies of the logit dynamics for the \ck\ game, starting in $\x$ and $\y$ respectively, coupled as described in Section~\ref{sec:coupling}. It is easy to check that, by Equation~(\ref{eq:up_prop}), the player selected for update, chooses strategy $0$ in both chain with probability at least $1/2$.

Finally, we bound the probability that after three steps the two coupled chains coalesce: it is at least as large as the probability that we choose three different players and all of them play strategy $0$ at their turn, i.e.
$$
\Prob{\mathbf{x},\mathbf{y}}{X_3 = Y_3} \geqslant \frac{1}{2} \cdot \frac{1}{3} \cdot \frac{1}{6} = \frac{1}{36}\,.
$$
Since this bound holds for every starting pair $(\x,\y)$, we have that the probability the two chains have not yet coalesced after $3t$ steps is
$$
\Prob{\mathbf{x},\mathbf{y}}{X_{3t} \neq Y_{3t}} \leqslant \left( 1 - \frac{1}{36} \right)^t \leqslant e^{-t/36}\,.
$$
The thesis follows from the Theorem~\ref{thm:coupling}.
\qed

%% file: trunk/twoplayers.tex
In this section we analyse the performance of the logit dynamics 
for $2\times 2$ coordination games (the same class studied in 
\cite{blumeGEB93}) and $2\times 2$ anti-coordination games.

% \subsection{Coordination games}
\paragraph{Coordination games.}
Coordination Games are two-player games 
in which the players have an advantage in selecting the same strategy. 
These games are often used to model the spread of a new technology \cite{youngTR00}:
two players have to decide whether to adopt or not a new technology. 
We assume that the players would prefer choosing the 
same technology as the other one and that choosing the new technology is at most as risky as choosing the old one.

We name $0$ the \texttt{NEW} strategy and $1$ the \texttt{OLD} strategy. 
The game is formally described by the following  payoff matrix
\begin{equation}
\label{eq:coorddef}
 \begin{array}{ | c || c | c | }
  \hline
   & 0 & 1 \\
  \hline \hline
  0 & (a, a) & (c, d) \\
  \hline
  1 & (d, c) & (b, b) \\
  \hline
\end{array}
\end{equation}
We assume that $a>d$ and $b>c$ (meaning that they prefer to coordinate)
and that $a-d\geq b-c$ (meaning that for each player strategy 0 is at most as risky as strategy 1). 
Notice that we do not make any assumption on the relation between $a$ and $b$. For convenience sake we name $$
\Delta := a-d \quad \mbox{ and } \quad\delta :=b-c\,.
$$
It is easy to see that this game is a potential game and the following function is 
an exact potential for it:
$$
 \Phi(0, 0)=\Delta  \qquad \Phi(0,1)=\Phi(1,0)=0  \qquad \Phi(1,1)=\delta.
$$
This game has two pure Nash equilibria: 
$(0, 0)$, where each player has utility $a$, and 
$(1, 1)$, where each player has utility $b$. 
As $d+c<a+b$, the social welfare is maximized at one of the two equilibria. 
% and the Price of Anarchy is equal to $\max\{b/a, a/b\}$.

We analyse the mixing time of the logit dynamics for $2\times 2$ coordination games and compute its stationary expected social welfare as a function of $\beta$.

\paragraph{Stationary expected social welfare.}
The logit dynamics for the coordination game defined by the payoffs in Table~\ref{eq:coorddef} establishes that,
from a profile $\x$, player $i$ selected for update plays according to the following probability distribution
(see Equation (\ref{eq:updateprob})):
$$
\begin{array}{ccccccc}
\sigma_i(0 \;|\; \x_{-i}=0) & = & \frac{1}{1 + e^{-\Delta\beta}}\,, & \qquad 
\sigma_i(1 \;|\; \x_{-i}=0) & = & \frac{1}{1 + e^{\Delta\beta}}\,,  \\[2mm]
\sigma_i(0 \;|\; \x_{-i}=1) & = & \frac{1}{1 + e^{\delta\beta}}\,,  & \qquad 
\sigma_i(1 \;|\; \x_{-i}=1) & = & \frac{1}{1 + e^{-\delta\beta}}\,.
\end{array}
$$
Next theorem bounds the stationary expected social 
welfare $\Expec{\pi}{W}$ obtained by the logit dynamics and gives conditions for which $\Expec{\pi}{W}$ is better than the social welfare $\welfare_N$ of the worst Nash equilibrium.
\begin{theorem}[Expected social welfare]
\label{thm:SFCoord}
The stationary expected social welfare $\Expec{\pi}{W}$
of the logit dynamics for the 
coordination game in Table~\ref{eq:coorddef} is  
$$
\Expec{\pi}{W} = 2 \cdot \frac{a+be^{- \left( \Delta-\delta \right)\beta}+(c+d)e^{-\Delta\beta}}{1 + e^{-\left( \Delta- \delta \right)\beta}+
					2e^{-\Delta\beta}}.
$$
Moreover, if $a \neq b$ then $\Expec{\pi}{W} \geq \welfare_N$ for $\beta$ sufficiently large.
\end{theorem}
\proof
The stationary distribution $\pi$ of the logit dynamics is
$$
             \pi(0, 0)  =  \frac{e^{\Delta\beta}}{Z(\beta)} \qquad
             \pi(1, 1)  =  \frac{e^{\delta\beta}}{Z(\beta)} \qquad
             \pi(0, 1) =  \pi(1, 0)  =  \frac{1}{Z(\beta)}
$$
where $Z(\beta) = e^{\Delta\beta} + e^{\delta\beta} + 2$.

Since $\Expec{\pi}{W}=2\cdot\Expec{\pi}{u_i}$, we compute the expected utility $\Expec{\pi}{u_i}$ of player  $i$ 
at the stationary distribution,
\begin{eqnarray*}
  \Expec{\pi}{u_i} &=& \sum_{\mathbf{x}\in\{0,1\}^2} u_i(\mathbf{x})\pi(\mathbf{x}) \\
                   &=& \frac{ae^{\Delta\beta}+be^{\delta\beta}+c+d}{e^{\Delta\beta}+e^{\delta\beta} + 2}\\
                   &=& \frac{a+be^{-(\Delta-\delta)\beta}+(c+d)e^{-\Delta\beta}}{1 + e^{-(\Delta-\delta)\beta}+2e^{-\Delta\beta}}\,.
\end{eqnarray*}
Thus, if $a > b$ and $\beta \geq \max\left\{0, \frac{1}{\Delta} \log \frac{2b - c - d}{a - b}\right\}$, we have
$$
 \Expec{\pi}{W} - \welfare_N = 2 \cdot \frac{a+be^{- \left( \Delta-\delta \right)\beta}+(c+d)e^{-\Delta\beta}}{1 + e^{-\left( \Delta- \delta \right)\beta}+2e^{-\Delta\beta}} - 2b = 2 \cdot \frac{(a - b)-(2b-c-d)e^{-\Delta\beta}}{1 + e^{-\left( \Delta- \delta \right)\beta}+2e^{-\Delta\beta}} \geq 0\,.
$$
Similarly, we obtain $\Expec{\pi}{W} - \welfare_N \geq 0$ if $b > a$ and $\beta \geq \max\left\{0, \frac{1}{\delta} \log \frac{2a - c - d}{b - a}\right\}$.
\qed

\paragraph{Mixing time.}
Now we study the mixing time of the logit dynamics for coordination games and we show that it is exponential in $\beta$ and in the minimum potential difference between adjacent profiles.
\begin{theorem}[Mixing Time]\label{theorem:coordmixing}
The mixing time of the logit dynamics for the coordination game of Table~\ref{eq:coorddef} is $\Theta\left(e^{\delta\beta}\right)$ for every $\beta \geq 0$.
\end{theorem}
\proof
\underline{Upper bound:}
We apply the Path Coupling technique 
(see Theorem~\ref{theorem:pathcoupling}) with the Hamming graph defined in~(\ref{eq:Hamming}) and all the edge-weights set to $1$. Let $\mathbf x$ and $\mathbf y$ be two profiles differing only for the player $j$ and consider the coupling defined in Section~\ref{sec:coupling} for this pair of profiles. Now we bound the expected distance of the two coupled chains after one step.

We denote by $b_i(\mathbf x,\mathbf y)$ the probability that both chains perform the same update given that player $i$ has been selected for strategy update. Clearly, $b_i(\mathbf x,\mathbf y)=1$ for $i=j$, while for $i\neq j$, we have
\begin{eqnarray*}
b_i(\mathbf{x},\mathbf{y}) & = &
\min\{\sigma_i(0 \;|\; \mathbf{x}), \, \sigma_i(0 \;|\; \mathbf{y}) \} +
\min\{\sigma_i(1 \;|\; \mathbf{x}), \, \sigma_i(1 \;|\; \mathbf{y}) \} \nonumber \\
& = & \frac{1}{1+e^{\Delta\beta}} + \frac{1}{1 + e^{\delta\beta}}\,.
\end{eqnarray*}
For sake of readability we set 
$$
p = \frac{1}{1+e^{\Delta\beta}} \qquad\text{and}\qquad q = \frac{1}{1 + e^{\delta\beta}}\,.
$$
and thus
$b_i(\mathbf{x},\mathbf{y})  = p+q$.
To compute 
$\Expec{\mathbf{x},\mathbf{y}}{\rho(X_1,Y_1)}$, 
we observe that the logit dynamics chooses player $j$ with probability 
$1/2$. In this case, as $b_j(\mathbf{x},\mathbf{y})=1$, the coupling updates both chains in the same way, resulting in $X_1=Y_1$. Similarly, player $i\ne j$ is chosen for strategy update with probability $1/2$.
In this case, with probability $b_i(\mathbf{x},\mathbf{y})$ the coupling performs the same update in both chains resulting in $\rho(X_1,Y_1)=1$. Instead with probability $1-b_i(\mathbf{x},\mathbf{y})$, the coupling performs different updates on the chains resulting in $\rho(X_1,Y_1)=2$. Therefore we have,
\begin{eqnarray*}
\Expec{\mathbf{x},\mathbf{y}}{\rho(X_1,Y_1)} 
& = & \frac{1}{2}b_i(\mathbf{x},\mathbf{y})+2\cdot\frac{1}{2}(1-b_i(\mathbf{x},\mathbf{y})) \\
& = & 1-\frac{1}{2} b_i(\mathbf{x},\mathbf{y}) 
=  1-\frac{1}{2}(p+q) \leqslant e^{-\frac{1}{2} (p+q)}\,.
\end{eqnarray*}
From Theorem~\ref{theorem:pathcoupling}, with $\alpha = \frac{1}{2} (p+q)$ and $\text{diam}(\Omega) = 2$, it follows that
$$
t_{\text{mix}}(\varepsilon) \leqslant \frac{2 \left(\log 2 + \log(1/\varepsilon)\right)}{p+q} = \frac{1}{p+q} \log \frac{4}{\varepsilon^2}\,.
$$

\smallskip\noindent
\underline{Lower bound:}
We use the \emph{relaxation time} bound (see Theorem~\ref{theorem:relaxation}).
The transition matrix of the logit dynamics is
$$
P =
\left(
\begin{array}{c|cccc}
& 00 & 01 & 10 & 11 \\
\hline
00 & 1-p     & p/2   & p/2   & 0 \\[1mm]
01 & \frac{1-p}{2} & \frac{p+q}{2} & 0             & \frac{1-q}{2} \\[1mm] 
10 & \frac{1-p}{2} & 0             & \frac{p+q}{2} & \frac{1-q}{2} \\[1mm]
11 & 0       & q/2   & q/2   & 1-q
\end{array}
\right)
$$
It is easy to see that the second largest eigenvalue of $P$ is 
$\lambda_\star = \frac{(1-p)+(1-q)}{2}$, hence the relaxation time is $t_{\text{rel}} = 1 / (1-\lambda_\star) = \frac{2}{p+q}$, and for the mixing time we have
\begin{eqnarray}\label{eq:lbmixcoord}
t_{\text{mix}}(\varepsilon) & \geqslant & (t_{\text{rel}} - 1)\log \frac{1}{2 \varepsilon} 
= \frac{2-(p+q)}{p + q} \log \frac{1}{2 \varepsilon} \nonumber \\
& \geqslant & \frac{1}{p + q} \log \frac{1}{2 \varepsilon}\,.
\end{eqnarray}
In the last inequality we used that $p$ and $q$ are both smaller than $1/2$.

Finally, the theorem follows by observing that
$$
\frac{1}{p + q} = \frac{1}{\frac{1}{1+e^{\Delta\beta}} + \frac{1}{1 + e^{\delta\beta}}}
= \Theta\left(e^{\delta \beta} \right)\,.
$$
\qed

Notice that, if we used the relaxation time to upper bound the mixing time (see Theorem~\ref{theorem:relaxation}) we would get a non-tight bound, hence in the above proof we had to resort to the path coupling for the upper bound.

% \subsection{Anti-coordination games}
% \label{sec:anti}
\paragraph{Anti-coordination games.}
Very similar results can be obtained for anti-coordination games.
These are two-player games in which the players 
have an advantage in selecting different strategies.
They model many settings where there is a common and 
exclusive resource: two players have to decide whether to use the resource or to drop it.
If they both try to use it, then a deadlock occurs and this is bad for both players. 
Usually, these games are described by a payoff matrix like the one in Table~\ref{eq:coorddef},
where we assume that $d>a$ and $c>b$
 and that $d-a\geq c-b$. 
Notice that Nash Equilibria of this game are \emph{unfair}, 
as one player has utility $\max\{c,d\}$ and  
the other $\min\{c,d\}$.

For the logit dynamics, we have that, for all $\beta$, the stationary expected social welfare
is worse than the one 
guaranteed by a Nash equilibrium. 
On the other hand, for sufficiently large $\beta$
we have that the expected utility of a player is always better than $\min\{c,d\}$:
that is, in the logit dynamics each player expects to gain more
than in the worst Nash equilibrium. 
Moreover, the stationary distribution is a fair equilibrium, since every player has the same expected utility.
As for the coordination games, the mixing time is exponential in $\beta$ and in the minimum potential difference between adjacent profiles.

%% file: trunk/orgame.tex
In this section we consider the following simple $n$-player
potential game that we here call \emph{OR game}. 
Every player has two strategies, say $\{0,1\}$, and  
each player pays the OR of the strategies of all players 
(including herself).
More formally, the utility function of player $i \in [n]$ is
$$
u_i(\mathbf{x}) =
\left\{
\begin{array}{cl}
 0, & \quad \mbox{if } \mathbf{x} = \boldsymbol{0}\,; \\
-1, & \quad \mbox{otherwise.}
\end{array}
\right.
$$
Notice that the OR game has $2^n - n$ Nash equilibria. 
The only profiles that are not Nash equilibria are the $n$ 
profiles with exactly one player playing $1$. 
Nash equilibrium $\boldsymbol{0}$ has social welfare $0$, 
while all the others have social welfare $-n$.

In Theorem~\ref{thm:SWOR} we show that the stationary expected social welfare is always better than the social welfare of the worst Nash equilibrium, and it is \emph{significantly} better for large $\beta$. Unfortunately, in Theorem~\ref{thm:lbOR} we show that if $\beta$ is large enough to guarantee a \emph{good} stationary expected social welfare, then the time needed to get close to the stationary distribution is exponential in $n$. Finally, in Theorem~\ref{thm:orupb} we give upper bounds on the mixing time showing that if $\beta$ is relatively small then the mixing time is polynomial in $n$, while for large $\beta$ the upper bound is exponential in $n$ and it is almost-tight with the lower bound. Despite the simplicity of the game, the analysis of the mixing time is far from trivial.

\begin{theorem}[Expected social welfare]
\label{thm:SWOR}
The stationary expected social welfare of the logit dynamics for the OR game is $\Expec{\pi}{W} = - \alpha n$ where $\alpha = \alpha(n,\beta) = \frac{(2^n - 1)e^{-\beta}}{1 + (2^n - 1)e^{-\beta}}$.
\end{theorem}
\proof
Observe that the OR game is a potential game with exact potential 
$\Phi$ where $\Phi(\boldsymbol{0}) = 0$ and $\Phi(\mathbf{x}) = -1$ for every $\mathbf{x} \neq \boldsymbol{0}$. Hence the stationary distribution is
$$
\pi(\mathbf{x}) =
\left\{
\begin{array}{ll}
1/Z, & \qquad \mbox{\rm if } \mathbf{x} = \boldsymbol{0}\,; \\[2mm]
e^{-\beta}/Z, & \qquad \mbox{\rm if } \mathbf{x} \neq \boldsymbol{0}\,;
\end{array}
\right.
$$
where the normalizing factor is $Z = 1 + (2^n - 1)e^{-\beta}$. 
The expected social welfare is thus
$$
\Expec{\pi}{W} 
	= \sum_{\mathbf{x}\in\{0,1\}^n} W(\mathbf{x})\pi(\mathbf{x}) 
	= -n\cdot\frac{(2^n - 1)e^{-\beta}}{1 + (2^n - 1)e^{-\beta}}\,.
$$
\qed

In the next theorem we show that the mixing time can be polynomial in $n$ only if $\beta \leqslant c \log n$ for some constant $c$.

\begin{theorem}[Lower bound on mixing time]
\label{thm:lbOR}
The mixing time of the logit dynamics for the OR game is
\begin{enumerate}
\item $\Omega(e^\beta)$ if $\beta < \log (2^n - 1)$;
\item $\Omega(2^n)$ if $\beta > \log (2^n - 1)$.
\end{enumerate}
\end{theorem}
%\begin{quote}
%\ideaproof Use the bottleneck ratio technique (See Theorem~\ref{theorem:bottleneck} in the Appendix). Use $S = \{ \boldsymbol{0} \}$ in the first case and the complementary set in the second case.
%\qed
%\end{quote}
\proof
Consider the set $S \subseteq\{0,1\}^n$ containing only the state $\boldsymbol{0} = (0, \dots, 0)$ and observe that $\pi(\boldsymbol{0}) \leqslant 1/2$ for $\beta \leqslant \log (2^n - 1)$. The bottleneck ratio is
$$
\Phi(\boldsymbol{0}) = \frac{1}{\pi(\boldsymbol{0})} \sum_{\mathbf{y} \in \{0,1\}^n} \pi(\boldsymbol{0}) P(\boldsymbol{0}, \mathbf{y}) = \sum_{\mathbf{y} \in \{0,1\}^n \,:\, |\mathbf{y}| = 1} P(\boldsymbol{0}, \mathbf{y}) = n \cdot \frac{1}{n} \frac{1}{1 + e^\beta}\,.
$$
Hence, by applying Theorem~\ref{theorem:bottleneck}, the mixing time is
$$
t_\text{mix} \geqslant \frac{1}{\Phi(\boldsymbol{0})} = 1 + e^{\beta}\,.
$$

\smallskip \noindent
If $\beta > \log (2^n - 1)$ instead we consider the set $R \subseteq\{0,1\}^n$ containing all states except state $\boldsymbol{0}$, and observe that
$$
\pi(R) = \frac{1}{Z} (2^n - 1) e^{-\beta} = \frac{(2^n - 1) e^{-\beta}}{1 + (2^n - 1) e^{-\beta}}\,.
$$
and $\pi(R) \leqslant 1/2$ for $\beta > \log (2^n - 1)$. It holds that
$$
Q(R,\overline{R}) = \sum_{\mathbf{x} \in R} \pi(\mathbf{x}) P(\mathbf{x}, \boldsymbol{0}) = \sum_{\mathbf{x} \in \{0,1\}^n \,:\, |\mathbf{x}| = 1} \pi(\mathbf{x}) P(\mathbf{x}, \boldsymbol{0}) = n \frac{e^{-\beta}}{Z} \frac{1}{n} \frac{1}{1+e^{-\beta}}\,.
$$
The bottleneck ratio is
$$
\Phi(R) = \frac{Q(R,\overline{R})}{\pi(R)} = \frac{Z}{(2^n-1)e^{-\beta}} \frac{e^{-\beta}}{Z} \frac{1}{1+e^{-\beta}} = \frac{1}{(2^n - 1)(1+e^{-\beta})} < \frac{1}{2^n - 1}\,.
$$
Hence, by applying Theorem~\ref{theorem:bottleneck}, the mixing time is
$$
t_{\text{mix}} \geqslant \frac{1}{\Phi(R)} > 2^n - 1\,.
$$
\qed

\smallskip \noindent
In the next theorem we give upper bounds on the mixing time depending on the value of $\beta$. The theorem shows that, if $\beta \leqslant c \log n$ for some constant $c$, the mixing time is effectively polynomial in $n$ with degree depending on $c$.
The use of the path coupling technique in the proof of the theorem requires a careful choice of the edge-weights.

\begin{theorem}[Upper bound on mixing time]
\label{thm:orupb}
The mixing time of the logit dynamics for the OR game is $\mathcal{O}(n^{5/2} 2^n)$ for every $\beta$. Moreover, for small values of $\beta$ the mixing time is
\begin{enumerate}
\item $\mathcal{O}(n \log n)$ if $\beta < (1-\varepsilon) \log n$, for an arbitrary small constant $\varepsilon > 0$;
\item $\mathcal{O}(n^{c + 3}\log n)$ if $\beta \leqslant c \log n$, where $c \geqslant 1$ is an arbitrary constant.
\end{enumerate}
\end{theorem}
\proof
We apply the path coupling technique (see Theorem~\ref{theorem:pathcoupling} in Section~\ref{apx:markov}) with the Hamming graph defined in~(\ref{eq:Hamming}). Let $\mathbf{x},\mathbf{y} \in \{ 0,1 \}^n$ be two profiles differing only at player $j \in [n]$ and, without loss of generality, let us assume $|\mathbf{x}| = k-1$ and $|\mathbf{y}| = k$ for some $k = 1, \dots, n$. We set the weight of edge $\{\mathbf{x},\mathbf{y}\}$ depending only on $k$, i.e. $\ell(\mathbf{x},\mathbf{y}) = \delta_k$ where $\delta_k \geqslant 1$ will be chosen later. Consider the coupling defined in Subsection~\ref{sec:coupling}.

Now we evaluate the expected distance after one step $\Expec{\mathbf{x},\mathbf{y}}{\rho(X_1,Y_1)}$ of the two coupled chains $(X_t,Y_t)$ starting at $(\mathbf{x},\mathbf{y})$. Let $i$ be the player chosen for the update. Observe that if $i=j$, i.e. if we update the player where $\mathbf{x}$ and $\mathbf{y}$ are different (this holds with probability $1/n$), then the distance after one step is zero, otherwise we distinguish four cases depending on the value of $k$.

\noindent \underline{Case $k = 1$}:
In this case profile $\mathbf{x}$ is all zeros and profile $\mathbf{y}$ has only one $1$ and the length of edge $\{\mathbf{x}, \mathbf{y}\}$ is $\ell(\mathbf{x}, \mathbf{y}) = \delta_1$. When choosing a player $i \neq j$ (this happens with probability $(n-1)/n$), at the next step the two chains will be at distance $\delta_1$ (if in both chains player $i$ chooses strategy $0$, and this holds with probability $\min\{ \sigma_i(0 \,|\, \mathbf{x}), \, \sigma_i(0 \,|\, \mathbf{y}) \}$), or at distance $\delta_2$ (if in both chains player $i$ chooses strategy $1$, and this holds with probability $\min\{ \sigma_i(1 \,|\, \mathbf{x}), \, \sigma_i(1 \,|\, \mathbf{y}) \}$), or at distance $\delta_1 + \delta_2$ (if player $i$ chooses strategy $0$ in chain $X_1$ and strategy $1$ in chain $Y_1$, and this holds with the remaining probability). Notice that, from the definition of the coupling, it will never happen that player $i$ chooses strategy $1$ in chain $X_1$ and strategy $0$ in chain $Y_1$, indeed we have that
\begin{equation}\label{eq:sigmask1}
\min\{ \sigma_i(0 \,|\, \mathbf{x}), \, \sigma_i(0 \,|\, \mathbf{y}) \} = \sigma_i(0 \,|\,\mathbf{y}) = \frac{1}{2}
\quad \mbox{ and } \quad
\min\{ \sigma_i(1 \,|\, \mathbf{x}), \, \sigma_i(1 \,|\, \mathbf{y}) \} = \sigma_i(1 \,|\,\mathbf{x}) = \frac{1}{1+e^\beta}\,.
\end{equation}
Hence the expected distance after one step is
\begin{eqnarray}
\Expec{\mathbf{x},\mathbf{y}}{\rho(X_1,Y_1)}
& = & \frac{n-1}{n} \left( \frac{1}{2} \delta_1 + \frac{1}{1+e^\beta} \delta_2 + \left(1 - \frac{1}{2} - \frac{1}{1 + e^\beta} \right) (\delta_1 + \delta_2) \right) \nonumber \\
& = & \frac{n-1}{n} \left( \frac{\delta_1}{1 + e^{-\beta}} + \frac{\delta_2}{2} \right)\,.
\end{eqnarray}

\noindent \underline{Case $k = 2$}: In this case we have $x_j = 0$ and $y_j = 1$, there is another player $h \in [n]\setminus \{j\}$ where $x_h = y_h = 1$, and for all the other players $i \in [n] \setminus \{ j,h \}$ it holds $x_i = y_i = 0$. Hence the length of edge $\{\mathbf{x}, \mathbf{y}\}$ is $\ell(\mathbf{x}, \mathbf{y}) = \delta_2$. 

When player $h$ is chosen (this holds with probability $1/n$) we have that $\sigma_h(s \,|\, \mathbf{x})$ and $\sigma_h(s \,|\, \mathbf{y})$ for $s = 0,1$ are the same as in (\ref{eq:sigmask1}). At the next step the two chains will be at distance $\delta_2$ (if player $h$ stays at strategy $1$ in both chains), or at distance $\delta_1$ (if player $h$ chooses strategy $0$ in both chains), or at distance $\delta_1 + \delta_2$ (if player $h$ stays at strategy $0$ in chain $X_1$ and chooses strategy $1$ in chain $Y_1$).

When a player $i \notin \{h,j\}$ is chosen (this holds with probability $(n-2)/n$) we have that $\sigma_i(0,\mathbf{x}) = \sigma_i(1,\mathbf{x}) = \sigma_i(0,\mathbf{y}) = \sigma_i(1,\mathbf{y}) = 1/2$. Thus in this case the two coupled chains always perform the same choice at player $i$, and at the next step they will be at distance $\delta_2$ (if player $i$ stays at strategy $0$ in both chains) or at distance $\delta_3$ (if player $i$ chooses strategy $1$ in both chains).

\noindent Hence the expected distance after one step is
\begin{eqnarray}
\Expec{\mathbf{x},\mathbf{y}}{\rho(X_1,Y_1)}
& = & \frac{1}{n} \left( \frac{1}{2} \delta_1 + \frac{1}{1+e^\beta} \delta_2 + \left(1 - \frac{1}{2} - \frac{1}{1 + e^\beta} \right) (\delta_1 + \delta_2) \right) + \frac{n-2}{n} \left( \frac{1}{2} \delta_2 + \frac{1}{2} \delta_3 \right) \nonumber \\
& = & \frac{1}{2n} \left( \frac{2}{1 + e^{-\beta}} \delta_1 + (n-1)\delta_2 + (n-2)\delta_3 \right)\,.
\end{eqnarray}

\noindent \underline{Case $3 \leqslant k \leqslant n-1$:} When a player $i \neq j$ is chosen such that $x_i = y_i = 1$ (this holds with probability $(k-1)/n$) then at the next step the two chains will be at distance $\delta_k$ (if $i$ stays at strategy $1$) or at distance $\delta_{k-1}$ (if $i$ moves to strategy $0$). When a player $i \neq j$ is chosen such that $x_i = y_i = 0$ (this holds with probability $(n-k)/n$) then at the next step the two chains will be at distance $\delta_k$ (if $i$ chooses to stay at strategy $0$) or at distance $\delta_{k+1}$ (if $i$ chooses to move to strategy $0$). Hence the expected distance after one step is
\begin{eqnarray}
\Expec{\mathbf{x},\mathbf{y}}{\rho(X_1,Y_1)}
& = & \frac{k-1}{n}\left( \frac{1}{2} \delta_k + \frac{1}{2} \delta_{k-1} \right) + \frac{n-k}{n} \left( \frac{1}{2}\delta_k + \frac{1}{2} \delta_{k+1} \right) \nonumber \\
& = & \frac{1}{2n}\left( (n-1)\delta_k + (k-1)\delta_{k-1} + (n-k)\delta_{k+1} \right)\,.
\end{eqnarray}

\noindent \underline{Case $k = n$:} When a player $i \neq j$ is chosen, then at the next step the two chains will be at distance $\delta_{n}$ or at distance $\delta_{n-1}$. Hence the expected distance after one step is
\begin{equation}
\Expec{\mathbf{x},\mathbf{y}}{\rho(X_1,Y_1)} = \frac{n-1}{n} \left(\frac{1}{2} \delta_n + \frac{1}{2}\delta_{n-1} \right) = \frac{n-1}{2n}(\delta_n + \delta_{n-1})\,.
\end{equation}

\noindent In order to apply Theorem~\ref{theorem:pathcoupling} we now have to show that it is possible to choose the edge weights $\delta_1, \dots, \delta_n$ and a parameter $\alpha > 0$ such that
\begin{equation}\label{eq:edineq}
\begin{array}{rcl}
\frac{n-1}{n} \left( \frac{\delta_1}{1 + e^{-\beta}} + \frac{\delta_2}{2} \right) & \leqslant & \delta_1 e^{-\alpha}\,, \\[2mm]
\frac{1}{2n} \left( \frac{2}{1 + e^{-\beta}} \delta_1 + (n-1)\delta_2 + (n-2)\delta_3 \right) & \leqslant & \delta_2 e^{-\alpha}\,, \\[2mm]
\frac{1}{2n}\left( (n-1)\delta_k + (k-1)\delta_{k-1} + (n-k)\delta_{k+1} \right) & \leqslant & \delta_k e^{-\alpha}\,, \qquad \mbox{for } k = 3, \dots, n-1\,, \\[2mm]
\frac{n-1}{2n}(\delta_n + \delta_{n-1}) & \leqslant & \delta_n e^{-\alpha}\,.
\end{array}
\end{equation}
For different values of $\beta$, we make different choices
for $\alpha$ and for the weights $\delta_k$.
For clarity's sake we split the proof in three different lemmas.
We denote by $\delta^{\max}$ the largest $\delta_k$.

In Lemma~\ref{lemma:ublargebeta} we show that
Inequalities~(\ref{eq:edineq}) are satisfied for every value of $\beta$ 
by choosing the weights as follows
$$
\delta_k =
\left\{
\begin{array}{ll}
\frac{1}{2} [(n-1)\delta_2 + 1], & \quad \mbox{\rm if \,} k = 1; \\[2mm]
\frac{n-k}{k} \delta_{k+1} + 1, & \quad \mbox{\rm if \,} 2 \leqslant k \leqslant n-1; \\[2mm]
1, & \quad \mbox{\rm if \,} k = n; \\[2mm]
\end{array}
\right.
$$
and by setting $\alpha = 1/(2 n \delta^{\max})$. From Corollary~\ref{cor:ublargebeta}, we have $\delta^{\max} = \mathcal{O}(\sqrt{n}2^n)$. Observe that the diameter of the Hamming graph is 
$\sum_{i = 1}^n \delta_i \leqslant n \delta^{\max}$, hence from Theorem~\ref{theorem:pathcoupling}
we obtain $t_{\text{mix}} = \mathcal{O}(n^{5/2}2^n)$.

In Lemma~\ref{lemma:ubsmallbeta} we show that, 
if $\beta < (1-\epsilon) \log n$ for an arbitrarily small constant $\varepsilon > 0$, 
Inequalities~(\ref{eq:edineq}) are satisfied, for sufficiently large $n$, by choosing weights 
$\delta_1 = n^{1-\varepsilon}$, 
$\delta_2 = 4/3, \delta_3 = \cdots = \delta_n = 1$, and $\alpha = 1/n$..
In this case the diameter is $\mathcal{O}(n)$ and, by Theorem~\ref{theorem:pathcoupling}, $t_{\text{mix}} = \mathcal{O}(n \log n)$.

%\noindent 
In Lemma~\ref{lemma:ubbetalog}
we show that, 
Inequalities~(\ref{eq:edineq}) are satisfied  
by choosing weights as follows
$$
\delta_k =
\left\{
\begin{array}{ll}
\frac{1 + e^{-\beta}}{2} \left[\frac{a_1}{b_1} \delta_2 + 1\right], & \quad \mbox{\rm if \,} k=1;\\[2mm]
\frac{a_k}{b_k} \delta_{k + 1} + 1, & \quad \mbox{\rm if \,} 2 \leqslant k \leqslant n-1; \\[2mm]
1, & \quad \mbox{\rm if \,} k = n;\\[2mm]
\end{array}
\right.
$$
where $a_1 = n - 1$ and $b_1 = n e^{-\beta} + 1$ and,
for every $k=2,\ldots,n-1$
$$a_k = (n - k)b_{k - 1} \qquad b_k=(n + 1)b_{k - 1} - (k - 1)a_{k - 1}\,;$$
and by setting $\alpha = 1/(2 n \delta^{\max})$.
From Corollary~\ref{cor:ubbetalog} it follows that, if $\beta \leqslant c\log n$ for a constant $c\in\mathbb{N}$, we have that $\delta_{\max} = \mathcal{O}(n^{c+2})$ and the diameter of the Hamming graph is $\mathcal{O}(n^{c+3})$. Thus, by Theorem~\ref{theorem:pathcoupling} it follows that $t_{\text{mix}} = \mathcal{O}(n^{c+3}\log n)$.
\qed

%% file: trunk/orgame-lemma.tex
In this section we prove the technical lemmas needed for completing
the proof of Theorem~\ref{thm:orupb}.
\begin{lemma}\label{lemma:ublargebeta}
Let $\delta_1, \dots, \delta_n$ be as follows
\begin{equation}\label{eq:defdeltaklargebeta}
\delta_k =
\left\{
\begin{array}{ll}
\frac{1}{2} [(n-1)\delta_2 + 1], & \quad \mbox{\rm if \,} k = 1; \\[2mm]
\frac{n-k}{k} \delta_{k+1} + 1, & \quad \mbox{\rm if \,} 2 \leqslant k \leqslant n-1;\\[2mm]
1, & \quad \mbox{\rm if \,} k=n;\\[2mm]
\end{array}
\right.
\end{equation}
and let $\alpha = 1/(2n\delta^{\max})$ where $\delta^{\max} = \max\{ \delta_k \;:\; k = 1, \dots, n \}$.
Then Inequalities~(\ref{eq:edineq}) are satisfied for every $\beta \geqslant 0$.
\end{lemma}
\proof
\noindent Observe that, for every $k=1, \dots, n$, the right-hand side of the $k$-th inequality in (\ref{eq:edineq}) is
\begin{equation}
 \label{eq:rhslargebeta}
 \delta_k e^{-\alpha} = \delta_k e^{-1/(2n\delta^{\max})} \geqslant \delta_k \left( 1 - \frac{1}{2n\delta^{\max}} \right) = \delta_k - \frac{\delta_k}{2n\delta^{\max}} \geqslant \delta_k - \frac{1}{2n}\,.
\end{equation}
Now we check that the left-hand side is at most $\delta_k - 1/(2n)$.

\smallskip \noindent \underline{First inequality ($k=1$):}
$\frac{n-1}{n} \left( \frac{\delta_1}{1 + e^{-\beta}} + \frac{\delta_2}{2} \right) \leqslant \delta_1 e^{-\alpha}$.

\noindent From the definition of $\delta_1$ in (\ref{eq:defdeltaklargebeta}) we have that
$$
\delta_2 = \frac{2 \delta_1 - 1}{n-1}\,.
$$
Hence the left-hand side is
$$
\frac{n-1}{n} \left( \frac{\delta_1}{1 + e^{-\beta}} + \frac{\delta_2}{2} \right)
\leqslant \frac{n-1}{n} \left( \delta_1 + \frac{\delta_2}{2} \right) \\
= \frac{n-1}{n} \left( \delta_1 + \frac{2\delta_1 - 1}{2(n-1)}\right)
= \frac{1}{2n} (2n \delta_1 - 1) = \delta_1 - \frac{1}{2n}\,.
$$

\smallskip \noindent \underline{Second inequality ($k=2$):}
$\frac{1}{2n} \left( \frac{2}{1 + e^{-\beta}} \delta_1 + (n-1)\delta_2 + (n-2)\delta_3 \right) \leqslant \delta_2 e^{-\alpha}$.

\noindent From the definition of $\delta_2$ in (\ref{eq:defdeltaklargebeta}) we have that
$$
\delta_3 = \frac{2}{n-2}(\delta_2 - 1)\,.
$$
Hence the left-hand side of the second inequality is
\begin{eqnarray*}
\frac{1}{2n} \left( \frac{2}{1 + e^{-\beta}} \delta_1 + (n-1)\delta_2 + (n-2)\delta_3 \right)
& \leqslant & \frac{1}{2n} \left( 2 \delta_1 + (n-1)\delta_2 + (n-2)\delta_3 \right) \\
& = & \frac{1}{2n} \left( (n-1) \delta_2 + 1 + (n-1)\delta_2 + 2(\delta_2 -1) \right) \\
& = & \frac{1}{2n}(2n \delta_2 - 1) = \delta_2 - \frac{1}{2n}\,.
\end{eqnarray*}

\smallskip \noindent \underline{Other inequalities ($k = 3, \dots, n-1$):} $\frac{1}{2n}\left( (n-1)\delta_k + (k-1)\delta_{k-1} + (n-k)\delta_{k+1} \right) \leqslant \delta_k e^{-\alpha}$.

\noindent From the definition of $\delta_k$ in (\ref{eq:defdeltaklargebeta}) we have that
$$
\delta_{k+1} = \frac{k}{n-k}(\delta_k - 1)\,.
$$
Hence the left-hand side is
\begin{eqnarray*}
\frac{1}{2n}\left( (n-1)\delta_k + (k-1)\delta_{k-1} + (n-k)\delta_{k+1} \right)
& = & \frac{1}{2n}\left( (n-1)\delta_k + (n-k+1)\delta_k + (k-1) + k \delta_k - k \right) \\
& = & \frac{1}{2n} (2n \delta_k - 1) = \delta_k - \frac{1}{2n}\,.
\end{eqnarray*}

\smallskip \noindent \underline{Last inequality ($k = n$):} $\frac{n-1}{2n}(\delta_n + \delta_{n-1}) \leqslant \delta_n e^{-\alpha}$.

\noindent Since $\delta_n = 1$ and $\delta_{n-1} = \frac{1}{n-1} \delta_n + 1 = \frac{n}{n-1}$, the left-hand side of the last inequality is
$$
\frac{n-1}{2n}(\delta_n + \delta_{n-1}) = \frac{n-1}{2n}(1 + \frac{n}{n-1}) = 1 - \frac{1}{2n}\,.
$$
\qed

\begin{lemma}\label{lemma:ubsmallbeta}
Let $\delta_1, \dots, \delta_n$ be as follows
$$
\delta_1=n^{1-\varepsilon}, \;\delta_2=4/3,\; \delta_3=\cdots=\delta_n = 1
$$
where $\varepsilon > 0$ is an arbitrary small constant and let $\alpha = 1/n$. 
Then Inequalities~(\ref{eq:edineq}) are satisfied for every 
$\beta \leqslant (1-\varepsilon) \log n$ and $n$ sufficiently large.
\end{lemma}
\proof We check that all the inequalities in (\ref{eq:edineq}) are satisfied.

\smallskip \noindent \underline{First inequality ($k=1$):}
$\frac{n-1}{n} \left( \frac{\delta_1}{1 + e^{-\beta}} + \frac{\delta_2}{2} \right) \leqslant \delta_1 e^{-\alpha}$.

\noindent For the left-hand side we have
\begin{eqnarray*}
\frac{n-1}{n} \left( \frac{\delta_1}{1 + e^{-\beta}} + \frac{\delta_2}{2} \right)
& = & \left(1 - \frac{1}{n} \right) \left( \frac{n^{1-\varepsilon}}{1 + e^{-\beta}} + \frac{2}{3} \right) \\
& \leqslant & \left(1 - \frac{1}{n} \right) \left( \frac{n^{1-\varepsilon}}{1 + \frac{1}{n^{1-\varepsilon}}} + \frac{2}{3} \right)
= \left(1 - \frac{1}{n} \right) \left( \frac{n^{2(1-\varepsilon)}}{n^{1-\varepsilon} + 1} + \frac{2}{3} \right) \\
& = & \left(1 - \frac{1}{n} \right) \left( \frac{(n^{1-\varepsilon} + 1)(n^{1-\varepsilon} - 1) + 1}{n^{1-\varepsilon} + 1} + \frac{2}{3} \right) \\
& = & \left(1 - \frac{1}{n} \right) \left( n^{1-\varepsilon} + \frac{1}{n^{1-\varepsilon} + 1} - \frac{1}{3} \right)\,.
\end{eqnarray*}
For the right-hand side we have
$$
\delta_1 e^{-\alpha} = n^{1-\varepsilon}e^{-1/n} \geqslant n^{1-\varepsilon}\left(1 - \frac{1}{n} \right)\,.
$$
Hence the left-hand side is smaller than the right-hand one (for $n$ sufficiently large).

\smallskip \noindent \underline{Second inequality ($k=2$):}
$\frac{1}{2n} \left( \frac{2}{1 + e^{-\beta}} \delta_1 + (n-1)\delta_2 + (n-2)\delta_3 \right) \leqslant \delta_2 e^{-\alpha}$.

\noindent For the left-hand side we have
\begin{eqnarray*}
\frac{1}{2n} \left( \frac{2}{1 + e^{-\beta}} \delta_1 + (n-1)\delta_2 + (n-2)\delta_3 \right)
& = & \frac{1}{2n} \left( \frac{2}{1 + e^{-\beta}} n^{1-\varepsilon} + (n-1)\frac{4}{3} + (n-2) \right)\\
& \leqslant & \frac{1}{2n} \left( 2 n^{1-\varepsilon} + \frac{7}{3} n \right) = \frac{7}{6} + \frac{1}{n^{\varepsilon}}\,.
\end{eqnarray*}
And for the right-hand side we have
$$
\delta_2 e^{-\alpha} = \frac{4}{3} e^{-1/n} \geqslant \frac{4}{3} \left(1 - \frac{1}{n} \right) \geqslant \frac{4}{3} - \frac{1}{n}\,.
$$
Hence the left-hand side is smaller than the right-hand one (for $n$ sufficiently large).

\smallskip \noindent \underline{Third inequality ($k = 3$):}
$\frac{1}{2n} \left( (n-1)\delta_3 + 2 \delta_2 + (n-3)\delta_4 \right) \leqslant \delta_3 e^{-\alpha}$.

\noindent For the left-hand side we have
\begin{eqnarray*}
\frac{1}{2n} \left( (n-1)\delta_3 + 2 \delta_2 + (n-3)\delta_4 \right) & = & \frac{1}{2n} \left( (n-1) + 2 \frac{4}{3} + (n-3) \right) \\
& = & \frac{1}{2n} \left( 2n - 3 \right) \leqslant \left( 1 - \frac{1}{n} \right)\,.
\end{eqnarray*}
And for the right-hand side we have
$$
\delta_3 e^{-\alpha} = e^{-1/n} \geqslant \left( 1 - \frac{1}{n} \right)\,.
$$
Hence the left-hand side is smaller than the right-hand one.

\smallskip \noindent \underline{Other inequalities ($k \geqslant 4$):}
$\frac{1}{2n}\left( (n-1)\delta_k + (k-1)\delta_{k-1} + (n-k)\delta_{k+1} \right) \leqslant \delta_k e^{-\alpha}$.

\noindent Since $\delta_k = \delta_{k-1} = \delta_{k+1} = 1$ the left-hand side is equal to $\frac{n-1}{n}$ and the right-hand side is $e^{-1/n} \geqslant \frac{n-1}{n}$.
\qed

\begin{lemma}\label{lemma:ubbetalog}
Let $\delta_1, \dots, \delta_n$ be as follows
\begin{equation}\label{eq:defdeltakbeta}
\delta_k =
\left\{
\begin{array}{ll}
\frac{1 + e^{-\beta}}{2} \left[\frac{a_1}{b_1} \delta_2 + 1\right], 
		& \quad \mbox{\rm if \,} k = 1;\\[2mm]
\frac{a_k}{b_k} \delta_{k + 1} + 1,
		& \quad \mbox{\rm if \,} 2 \leqslant k \leqslant n-1;\\[2mm]
1, 
		& \quad \mbox{\rm if \,} k = n;\\[2mm]
\end{array}
\right.
\end{equation}
where $a_1=n-1$ and $b_1=n e^{-\beta} + 1$ and for every $k = 2, \dots, n - 1$
$$
 a_k = (n - k)b_{k - 1} \qquad \text{and} \qquad b_k=(n + 1)b_{k - 1} - (k - 1)a_{k - 1}\,,
$$
and let $\alpha = 1/(2n\delta^{\max})$ where 
$\delta^{\max} = \max\{ \delta_k \;:\; k = 1, \dots, n \}$.
Then Inequalities~(\ref{eq:edineq}) are satisfied for every $\beta \geqslant 0$.
\end{lemma}
Before to prove the Lemma~\ref{lemma:ubbetalog} we do the following observation.
\begin{obs}
 \label{obs:b_property}
 Let $b_k$ defined as in the Lemma~\ref{lemma:ubbetalog}. Then, for every $k \geqslant 2$, it holds that $b_k \geqslant k b_{k - 1}$.
\end{obs}
\proof  We proceed by induction on $k$. 
The base case $k=2$ follows from
$$
b_2 = 
	(n + 1) (n e^{-\beta} + 1) - (n - 1) = 
	(n + 1) n e^{-\beta} + 2 > 2(n e^{-\beta} + 1) = 
	2 b_1\,.
$$
Now suppose the claim holds for $k-1$, 
that is $b_{k - 1} \geqslant (k - 1)b_{k - 2}$. Then
\begin{eqnarray*}
 b_k & = & (n + 1) b_{k - 1} - (k - 1) a_{k - 1}\\
     & = & (n + 1) b_{k - 1} - (k - 1) (n - k + 1) b_{k - 2}\\
     & \geqslant & \left[(n + 1) - (n - k + 1)\right] b_{k - 1} = k b_{k - 1}\,.
\end{eqnarray*}
\qed

\noindent \textit{Proof (Lemma~\ref{lemma:ubbetalog}).}
\noindent
Observe that, as in Equation~(\ref{eq:rhslargebeta}), for every $k=1, \dots, n$, the right-hand side of the $k$-th inequality in (\ref{eq:edineq}) is
$$
\delta_k e^{-\alpha} \geqslant \delta_k - \frac{1}{2n}\,.
$$
Now we check that the left-hand side is at most $\delta_k - 1/(2n)$.

\smallskip \noindent \underline{First inequality ($k=1$):}
$\frac{n-1}{n} \left( \frac{\delta_1}{1 + e^{-\beta}} + \frac{\delta_2}{2} \right) \leqslant \delta_1 e^{-\alpha}$.

\noindent From the definition of $\delta_1$ in (\ref{eq:defdeltakbeta}) we have that
$$
\delta_{2} = \frac{n e^{-\beta} + 1}{n - 1}\left(\frac{2\delta_1}{1 + e^{-\beta}} - 1\right)\,.
$$
Hence the left-hand side is
\begin{eqnarray*}
\frac{n-1}{n} \left( \frac{\delta_1}{1 + e^{-\beta}} + \frac{\delta_2}{2} \right)
& = & \frac{n-1}{n} \left[ \frac{\delta_1}{1 + e^{-\beta}} + \frac{ne^{-\beta} + 1}{n - 1} \left( \frac{\delta_1}{1 + e^{-\beta}} - \frac{1}{2} \right) \right] \\
& = & \frac{n-1}{n} \frac{\delta_1}{1 + e^{-\beta}} \left( 1 + \frac{ne^{-\beta} + 1}{n - 1} \right) - \frac{ne^{-\beta} + 1}{2n} \\
& \leqslant & \delta_1 - \frac{1}{2n}\,.
\end{eqnarray*}

\smallskip \noindent \underline{Second inequality ($k=2$):}
$\frac{1}{2n} \left( \frac{2}{1 + e^{-\beta}} \delta_1 + (n-1)\delta_2 + (n-2)\delta_3 \right) \leqslant \delta_2 e^{-\alpha}$.

\noindent From the definition of $\delta_2$ in (\ref{eq:defdeltakbeta}) we have that
$$
\delta_{3} = \frac{b_2}{a_2} (\delta_2 - 1) = \frac{(n + 1)b_1 - a_1}{(n - 2) b_1} (\delta_2 - 1)\,.
$$
Hence the left-hand side is
\begin{eqnarray*}
\frac{1}{2n} \left( \frac{2}{1 + e^{-\beta}} \delta_1 + (n-1)\delta_2 + (n-2)\delta_3 \right)
& = & \frac{1}{2n} \left[ \left(\frac{a_1}{b_1} \delta_2 + 1\right) + (n-1)\delta_2 + \frac{(n + 1)b_1 - a_1}{b_1} (\delta_2 - 1) \right] \\
& = & \delta_2 - \frac{1}{2n} \frac{n b_1 - a_1}{b_1} = \delta_2 - \frac{1}{2n} \left( n - \frac{n - 1}{n e^{-\beta} + 1} \right) \\
& \leqslant & \delta_2 - \frac{1}{2n}\,.
\end{eqnarray*}

\smallskip \noindent \underline{Other inequalities ($k = 3, \dots, n-1$):}
$\frac{1}{2n}\left( (n-1)\delta_k + (k-1)\delta_{k-1} + (n-k)\delta_{k+1} \right) \leqslant \delta_k e^{-\alpha}$.

\noindent From the definition of $\delta_k$ in (\ref{eq:defdeltakbeta}) we have that
$$
\delta_{k + 1} = \frac{b_k}{a_k} (\delta_k - 1) = \frac{(n + 1)b_{k - 1} - (k - 1) a_{k - 1}}{(n - k)b_{k - 1}} (\delta_k - 1)\,.
$$
Hence the left-hand side is
\begin{eqnarray*}
\frac{1}{2n}\left( (n-1)\delta_k + (k-1)\delta_{k-1} + (n-k)\delta_{k+1} \right)
& = & \frac{1}{2n} \left[ (n-1)\delta_k + (k - 1) \left(\frac{a_{k - 1}}{b_{k - 1}} \delta_k + 1 \right) \right.\\
& + & \left. \frac{(n + 1)b_{k - 1} - (k - 1) a_{k - 1}}{b_{k - 1}} (\delta_k - 1) \right] \\
& = & \delta_k - \frac{1}{2n} \frac{(n - k + 2) b_{k - 1} - (k - 1) a_{k - 1}}{b_{k - 1}} \\
& = & \delta_k - \frac{1}{2n} \left( (n - k + 2) - (k - 1) (n - k + 1)\frac{b_{k - 2}}{b_{k - 1}} \right) \\
& \leqslant & \delta_2 - \frac{1}{2n}\,.
\end{eqnarray*}
where the inequality follows from the Observation~\ref{obs:b_property}.

\smallskip \noindent \underline{Last inequality ($k = n$):}
$\frac{n-1}{2n}(\delta_n + \delta_{n-1}) \leqslant \delta_n e^{-\alpha}$.

\noindent Since $\delta_n = 1$ and $\delta_{n-1} = \frac{a_{n - 1}}{b_{n - 1}} \delta_n + 1 = \frac{a_{n - 1}}{b_{n - 1}} + 1$, the left-hand side of the last inequality is
\begin{eqnarray*}
\frac{n-1}{2n}(\delta_n + \delta_{n-1})
& = & \frac{n - 1}{2n} \left(2 + \frac{a_{n - 1}}{b_{n - 1}}\right) = \frac{n - 1}{2n} \left( 2 + \frac{b_{n - 2}}{b_{n - 1}} \right) \\
& \leqslant & \frac{n - 1}{2n} \left( 2 + \frac{1}{n - 1} \right) = 1 - \frac{1}{2n}\,.
\end{eqnarray*}
where the inequality follows from the Observation~\ref{obs:b_property}.
\qed

In order to apply the path coupling theorem, we need to bound $\delta_{\max}$: the next observation will represent the main tool to achieve this goal.
\begin{obs}\label{obs:maxweight}
Let $\delta_1, \dots, \delta_n$ be defined recursively as follows: 
$\delta_n = 1$ and
$$ \delta_{k} = \gamma_k \delta_{k+1} + 1\,, $$
where $\gamma_k>0$ for every $k=1, \dots, n-1$. 
Let $\delta^{\max} = \max\{ \delta_k \;:\; k = 1, \dots, n \}$.
Then
$$
\delta^{\max} \leqslant n \max\left\{\prod_{i=h}^j \gamma_i \;:\; 1 \leqslant h \leqslant j \leqslant n-1 \right\}\,.
$$
\end{obs}
\proof
The observation follows from the fact that,
for $k = 1, \dots, n-1$, we have
$$
\delta_k = 1 + \sum_{j=k}^{n-1} \prod_{i=k}^j \gamma_i\,.
$$
\qed

\begin{cor}\label{cor:ublargebeta}
Let $\delta_1, \dots, \delta_n$ be defined as in Lemma~\ref{lemma:ublargebeta}. Then
$
\delta^{\max} \leqslant c \sqrt{n} 2^n
$
for a suitable constant $c$.
\end{cor}
\proof
From Observation~\ref{obs:maxweight} and 
the definition of $\delta_1, \dots, \delta_n$,
it holds that
\begin{eqnarray*}
\delta^{\max} & \leqslant & n \max\left\{\prod_{i=h}^j \frac{n-i}{i} \;:\; 1 \leqslant h \leqslant j \leqslant n \right\} \\
& \leqslant & n \prod_{i=1}^{\lfloor n/2 \rfloor} \frac{n-i}{i} \leqslant n \binom{n}{\lfloor n/2 \rfloor} \leqslant c \sqrt{n} 2^n\,.
\end{eqnarray*}
for a suitable constant $c$.
\qed

In order to bound $\delta_{\max}$ when $\delta_1, \dots, \delta_n$ are defined as in Lemma~\ref{lemma:ubbetalog} and 
$\beta \leqslant c \log n$ for a constant $c \in \mathbb{N}$, we define
\begin{equation}
\label{eq:gamma_k}
 \gamma_k = \frac{a_k}{b_k} = \frac{p_k e^{-\beta} + l_k}{q_k e^{-\beta} + r_k}\,.
\end{equation}
You can check that $p_1 = 0$, $q_1 = n$ and
$$
  p_k = (n - k)q_{k - 1} \qquad q_k = (n + 1) q_{k - 1} - (k - 1) p_{k - 1}\,;
$$
we notice that $p_k = (n + 1)q_{k - 1} - (k + 1) q_{k - 1} \leq q_k$ for every $k$.
We can also prove the following simple observation about $q_k$.
\begin{obs}
\label{eq:q_k_constant}
For every $k \geq 1$ constant, we have $q_k \geq 2^{-k} n^k$.
\end{obs}
\proof
We proceed by induction on $k$, with the base $k=1$ being obvious.
Suppose the claim holds for $k-1$, that is $q_{k - 1} \geq 2^{-(k - 1)} n^{k - 1}$, then
$$
   q_k = (n + 1)q_{k - 1} - (k - 1)p_{k - 1} \geq \frac{n}{2} q_{k - 1} \geq 2^{-k} n^k\,.
$$
\qed

Moreover, you can check that $l_1 = n - 1$, $r_1 = 1$ and
$$
 l_k = (n - k) r_{k - 1} \qquad r_k = (n + 1) r_{k - 1} - (k - 1) l_{k - 1}\,;
$$
we notice that above recursion gives $l_k = (n - k) (k - 1)!$ and $r_k = k!$.
Next lemma bounds $\gamma_k$ defined in Equation~\ref{eq:gamma_k}.
\begin{lemma}\label{lemma:gamma}
Let $\delta_1, \dots, \delta_n$ be defined as in Lemma~\ref{lemma:ubbetalog}, $\gamma_k$ defined as in Equation~(\ref{eq:gamma_k})
and 
$\beta \leqslant c \log n$ for a constant $c \in \mathbb{N}$. Then, for sufficiently large $n$, it holds that
\begin{equation*}
\left\{
\begin{array}{cccl}
 \gamma_k & < & n & \quad \forall \; k; \\[2mm]
 \gamma_k & < & 1 & \quad \mbox{\rm if } k > c + 2;\\[2mm]
 \gamma_{c + 2} & = & \mathcal{O}(1).
\end{array}
\right.
\end{equation*}
\end{lemma}
\proof
Since $p_k \leq q_k$, then $(n q_k - p_k)e^{-\beta} > 0$; instead, $l_k - n r_k = (k - 1)! (n - k - nk) < 0$. Hence we have for every $k$
 $$
  \gamma_k - n = \frac{p_k e^{-\beta} + l_k}{q_k e^{-\beta} + r_k} - n = \frac{(l_k - n r_k) - (n q_k - p_k)e^{-\beta}}{q_k e^{-\beta} + r_k} < 0\,.
 $$

Inductively, we show that for every $k \geq c + 3$, we have $\gamma_k < 1$. Set $k = c + 3$: $c$ is a constant, thus Observation~\ref{eq:q_k_constant} holds for $k - 1$; hence and since $e^{-\beta} \geqslant n^{-c}$, we have that
 $$
  (q_{c + 3} - p_{c + 3})e^{-\beta} = [(n + 1) q_{c + 2} - (c + 2) p_{c + 2} - (n - c - 3) q_{c + 2}]e^{-\beta} \geqslant 2 q_{c + 2}e^{-\beta} \geqslant 2^{-(c + 1)} n^{2}.
 $$
Instead, $l_{c + 3} - r_{c + 3} = (c+2)!(n - 2c - 6) \leqslant (c + 2)! \cdot n$. Thus,
 $$
  \gamma_{c+3} - 1 = \frac{(l_{c+3} - r_{c+3}) - (q_{c+3} - p_{c+3})e^{-\beta}}{q_{c+3} e^{-\beta} + r_{c+3}} \leq \frac{(c + 2)! \cdot n - 2^{-(c + 1)} n^{2}}{q_{c+3} e^{-\beta} + r_{c+3}} < 0\,,
 $$
for $n$ sufficiently large. 
Now, suppose that $\gamma_{k - 1} < 1$; then, we have
 $$
  \gamma_k - 1 = \frac{a_k - b_k}{b_k} = \frac{(k - 1) a_{k - 1} - (k + 1) b_{k - 1}}{b_k} < 0\,,
 $$
where $a_{k - 1} < b_{k - 1}$ is implied by the inductive hypothesis.

In order to complete the proof, we need to show that $\gamma_{c + 2} = \mathcal{O}(1)$.
Similarly to the case $k = c+3$, we obtain $(q_{c + 2} - p_{c + 2})e^{-\beta} \geqslant 2^{-c} n$ and $l_{c + 2} - r_{c + 2} \leqslant (c + 1)! \cdot n$. Hence,
$$
  \gamma_{c + 2} \leqslant \frac{p_{c+2} + r_{c+2} + (c+1)! \cdot n}{p_{c+2} + r_{c+2} + 2^{-c}n} \leq (c + 1)! \cdot 2^c = \mathcal{O}(1)\,.
 $$
\qed

% Finally, we are ready to bound $\delta_{\max}$ in Lemma~\ref{lemma:ubbetalog}.
\begin{cor}\label{cor:ubbetalog}
Let $\delta_1, \dots, \delta_n$ and $c$ be defined as in Lemma~\ref{lemma:ubbetalog}. Then
$
\delta^{\max} = \mathcal{O}(n^{c + 2}).
$
\end{cor}
\proof
From Observation~\ref{obs:maxweight}, 
Lemma~\ref{lemma:gamma} and the definition of 
$\delta_1,\dots,\delta_n$ it follows that
\begin{eqnarray*}
\delta^{\max} & \leqslant & n \max\left\{\prod_{i=h}^j \frac{a_i}{b_i} \;:\; 1 \leqslant h \leqslant j \leqslant n \right\} \\
& \leqslant & n \prod_{i=1}^{c + 2} \frac{a_i}{b_i} = \mathcal{O}(n^{c + 2}).
\end{eqnarray*}
\qed

%% file: trunk/xor.tex
In this section we analyze the
logit dynamics for another simple $n$-player game, the {\em XOR game}.
The XOR game is a symmetric $n$-player game in which each
player has two strategies, denoted by $0$ and $1$, and
each player pays the XOR of the strategies of all players
(including herself).
More formally, for each $i\in[n]$, the utility function
$u_i(\cdot)$ is defined as follows
$$u_i(\x)=\begin{cases}
	-1, & \text{if } \x \text{ has an odd number of }$1$\text{'s;}\cr
	\phantom{-}0, & \text{if } \x \text{ has an even number of }$1$\text{'s.}\cr
	  \end{cases}
$$
Notice that the XOR game has $2^{n - 1}$ Nash equilibria, namely all profiles with an even number of players playing strategy 1.
Nash equilibria have social welfare $0$ and profiles not in equilibria have social welfare $-n$.
Observe that the XOR game is a potential game with exact potential
$\Phi$ where $\Phi(\x) = u_i(\x)$ for every $\x$ and every $i\in[n]$.
Hence, the stationary distribution is
$$
\pi(\mathbf{x}) = \begin{cases}
                   e^{-\beta}/Z, & \text{if } \x \text{ has an odd number of }$1$\text{'s;}\cr
                   1/Z, & \text{if } \x \text{ has an even number of }$1$\text{'s;}\cr
                  \end{cases}
$$
where the normalizing factor is $Z = 2^{n - 1}(1 + e^{-\beta})$.

Even if this game looks similar to the OR game, it exhibits a different behavior.
Theorem~\ref{thm:SWXOR} gives the
stationary expected social welfare of the XOR game and
we can see that, as $\beta$ increases,
the expected social welfare tends from below to the social welfare at
the Nash equilibria.
In contrast the expected social welfare of the OR game is better than
the worst Nash equilibrium for all values of $\beta$.
Moreover, in Theorem~\ref{thm:lbXOR} and Theorem~\ref{thm:xorupb}
we show that the mixing time for the XOR game
is polynomial in $n$ and exponential in $\beta$,
whereas the mixing time for the OR game can be bounded
independently from $\beta$.
%Finally, we mention that the upper bound on the mixing time of the XOR game
%was obtained using coupling whereas for the OR game the path coupling
%technique was sufficient.

\begin{theorem}[Expected social welfare]
\label{thm:SWXOR}
The stationary expected social welfare of the logit dynamics for the XOR game is $\Expec{\pi}{W} = - \frac{n}{1 + e^\beta}$.
\end{theorem}
\proof
The expected social welfare is
$$
\Expec{\pi}{W}
	= \sum_{\mathbf{x}\in\{0,1\}^n} W(\mathbf{x})\pi(\mathbf{x})
	= -n\cdot\frac{2^{n - 1}e^{-\beta}}{2^{n - 1}(1 + e^{-\beta})} = - \frac{n}{1 + e^\beta}.
$$
\qed

The next theorem shows that the mixing time is exponential
in $\beta$ for every $\beta > 0$.
\begin{theorem}[Lower bound on mixing time]
\label{thm:lbXOR}
The mixing time of the logit dynamics for the XOR game is $\Omega(e^\beta)$.
\end{theorem}
\proof
Consider the set $S \subseteq\{0,1\}^n$ containing only the state $\boldsymbol{0} = (0, \dots, 0)$. Observe that $\pi(\boldsymbol{0}) \leqslant 1/2$. The bottleneck ratio is
$$
B(\boldsymbol{0}) = \frac{1}{\pi(\boldsymbol{0})} \sum_{\y \in \{0,1\}^n} \pi(\boldsymbol{0}) P(\boldsymbol{0}, \y) = \sum_{\y \in \{0,1\}^n \,:\, |\y| = 1} P(\boldsymbol{0}, \y) = n \cdot \frac{1}{n}\cdot\frac{1}{1 + e^\beta}\,.
$$
Hence, by applying Theorem~\ref{theorem:bottleneck}, the mixing time is
$$
t_\text{mix} \geqslant \frac{1}{B(\boldsymbol{0})} = 1 + e^{\beta}\,.
$$
\qed

Finally, in the next theorem we give an almost matching upper bound to the mixing time.
\begin{theorem}[Upper bound on mixing time]
\label{thm:xorupb}
The mixing time of the logit dynamics for the OR game is $\mathcal{O}(n^3 e^\beta)$.
\end{theorem}
The theorem is proved using coupling (see Theorem~\ref{thm:coupling}) and
proof is presented in the next sections.
Specifically,
we use the coupling described in Section~\ref{sec:coupling};
in Section~\ref{sec:xor_even} we show that if the coupled chains are
at even distance then distance does not increase after one step of the coupling;
in Section~\ref{sec:xor_odd} we show that if the coupled chains are
at odd distance then they get closer distance with probability independent from $\beta$;
finally, in Section~\ref{sec:xor_coalesce} we bound the expected time needed
by the two chains to coalesce and use Theorem~\ref{thm:coupling} to derive
an upper bound for the mixing time.

\subsection{Even Hamming distance}
\label{sec:xor_even}
Let $X_t$ and $Y_t$ be two chains coupled as described in Section~\ref{sec:coupling}. Suppose that $X_t=\mathbf{x}$, $Y_t=\mathbf{y}$, and
$\hamming(\x,\y)=2\ell$, for $\ell>0$.
In this case, $u_i(\x)=u_i(\y)=b$ for all $i\in[n]$ and some $b\in\{-1,0\}$.

Let $i$ be the index selected for update and let us distinguish two cases.
In the first case $x_i=y_i$ and we have
$$
	u_i(\mathbf{x}_{-i},0)=u_i(\mathbf{y}_{-i},0)
		\quad\textrm{and}\quad
	u_i(\mathbf{x}_{-i},1)=u_i(\mathbf{y}_{-i},1)$$
and thus
$$
	\sigma_i(0 \mid \mathbf{x})=\sigma_i(0 \mid \mathbf{y})
		\quad\textrm{and}\quad
	\sigma_i(1 \mid \mathbf{x})=\sigma_i(1 \mid \mathbf{y}).$$
Therefore the coupling always update the strategy of player $i$ in the same
way in the two chains and thus
$\hamming(X_{t+1},Y_{t+1})=2\ell$.

In the second case we have $x_i\ne y_i$ and we assume,
without loss of generality, that $x_i=0$ and $y_i=1$.
We observe that, for $b \in \{-1,0\}$,
$$
	u_i(\mathbf{x}_{-i},0)=u_i(\mathbf{y}_{-i},1)=b
		\quad\textrm{and}\quad
	u_i(\mathbf{y}_{-i},0)=u_i(\mathbf{x}_{-i},1)=-(1+b).
$$
Therefore we have
$$
\sigma_i(0 \mid \mathbf{x})=\sigma_i(1 \mid \mathbf{y})=\frac{1}{1+e^{-(1+2b)\beta}}
		\quad\textrm{and}\quad
\sigma_i(1 \mid \mathbf{x})=\sigma_i(0 \mid \mathbf{y})=\frac{1}{1+e^{(1+2b)\beta}}
$$
and thus we have three possible updates for the strategy of
player $i$:
\begin{enumerate}
\item both chains update to $0$
	(and thus $\hamming(X_{t+1},Y_{t+1})=2\ell-1$)
	with probability
	$$\min\left\{
		\frac{1}{1+e^{(1+2b)\beta}},
		\frac{1}{1+e^{-(1+2b)\beta}}
		\right\}=\frac{1}{1+e^\beta};$$
\item both chains update to $1$
	(and thus $\hamming(X_{t+1},Y_{t+1})=2\ell-1$)
	with probability
	$$\min\left\{
		\frac{1}{1+e^{(1+2b)\beta}},
		\frac{1}{1+e^{-(1+2b)\beta}}
		\right\}=\frac{1}{1+e^\beta};$$

\item chain $X$ and $Y$ choose two different
strategies for updating the strategy of player $i$
		(and thus $\hamming(X_{t+1},Y_{t+1})=2\ell$)
	with probability
	$$1-\frac{2}{1+e^{\beta}}.$$
\end{enumerate}
The following lemma summarizes the above observations.
\begin{lemma}\label{le:evenl}
Suppose that $\hamming(X_t,Y_t)=2\ell$, for $\ell>0$.
Then
$$
	\hamming(X_{t+1},Y_{t+1})=
	\begin{cases}
2\ell-1,&\textrm{with probability }   \frac{2\ell}{n}\cdot\frac{2}{1+e^\beta};\\
\\
2\ell,  &\textrm{with probability } 1-\frac{2\ell}{n}\cdot\frac{2}{1+e^\beta}.\\
	\end{cases}
$$
\end{lemma}

\subsection{Odd Hamming distance}
\label{sec:xor_odd}
Let $X_t$ and $Y_t$ be two chains coupled as described in Section~\ref{sec:coupling}.
Suppose that $X_t=\mathbf{x}$, $Y_t=\mathbf{y}$, and $\hamming(\x,\y)=2\ell-1$, for $\ell>0$.
In this case we have $u_i(\x)=b$ and $u_i(\y)=-(1+b)$ for some $b\in\{-1,0\}$.
Let $i$ be the index selected for update and let us distinguish two cases.

In the case in which $x_i=y_i=c$ for some $c\in\{0,1\}$, we have
$$
u_i(\mathbf{x}_{-i},c)=u_i(\mathbf{y}_{-i},1-c)=b
	\quad\textrm{and}\quad
u_i(\mathbf{x}_{-i},1-c)=u_i(\mathbf{y}_{-i},c)=-(1+b).
$$
Therefore
$$
\sigma_i(c \mid \mathbf{x})=\sigma_i(1-c \mid \mathbf{y})=\frac{1}{1+e^{-(1+2b)\beta}}
	\quad\textrm{and}\quad
\sigma_i(1-c \mid \mathbf{x})=\sigma_i(c \mid \mathbf{y})=\frac{1}{1+e^{ (1+2b)\beta}}
$$
and thus we have three possible updates:
\begin{enumerate}
\item  both chains update to $c$ (and thus $\hamming(X_{t+1},Y_{t+1})=2\ell-1$)
	with probability
	$$\min\left\{
		\frac{1}{1+e^{-(1+2b)\beta}},
		\frac{1}{1+e^{ (1+2b)\beta}}
		\right\}= \frac{1}{1+e^{\beta}};$$
\item both chains update to $1-c$
	(and thus $\hamming(X_{t+1},Y_{t+1})=2\ell-1$)
	with probability
	$$\min\left\{
		\frac{1}{1+e^{-(1+2b)\beta}},
		\frac{1}{1+e^{ (1+2b)\beta}}
		\right\}= \frac{1}{1+e^{\beta}};$$
\item chains $X$ and $Y$ choose two different strategies for updating
	the strategy player $i$
	(and thus $\hamming(X_{t+1},Y_{t+1})=2\ell$)
	with probability $1-\frac{2}{1+e^{\beta}}$.
\end{enumerate}

\smallskip\noindent
In the second case we have $x_i\ne y_i$ and we assume,
without loss of generality, that $x_i=0$ and $y_i=1$.
We observe that
$$
	u_i(\mathbf{x}_{-i},0)=u_i(\mathbf{y}_{-i},0)=b
	\quad\textrm{and}\quad
	u_i(\mathbf{x}_{-i},1)=u_i(\mathbf{y}_{-i},1)=-(1+b).
$$
Therefore we have
$$
	\sigma_i(0 \mid \mathbf{x})=\sigma_i(0 \mid \mathbf{y})
	\quad\textrm{and}\quad
	\sigma_i(1 \mid \mathbf{x})=\sigma_i(1 \mid \mathbf{y})
$$
and thus in this case $\hamming(X_{t+1},Y_{t+1})=2\ell-2$.

The following lemma summarizes the above observations.
\begin{lemma}\label{le:oddl}
Suppose that $\hamming(X_t,Y_t)=2\ell-1$, for $\ell>0$.
Then
$$
	\hamming(X_{t+1},Y_{t+1})=
	\begin{cases}
	2\ell-2,  & \textrm{with probability }  \frac{2\ell-1}{n}; \\
\\
	2\ell-1,  & \textrm{with probability }  \frac{n-2\ell+1}{n}
					\frac{2}{1+e^\beta};\\
\\
	2\ell  ,  & \textrm{with probability }  \frac{n-2\ell+1}{n}
			\left( 1- \frac{2}{1+e^\beta}\right). \\
	\end{cases}
$$
\end{lemma}

\subsection{Time to coalesce}
\label{sec:xor_coalesce}
We denote with $\tau_k$ the random variable indicating the first time at which
the two coupled chains have distance $k$. More precisely,
$$\tau_k=\min\{t: \hamming(X_t,Y_t)=k\}.$$
Therefore, $\tc=\tau_0$ is the time needed for the two chains to coalesce.
We next give a bound on the expected time
$\Expec{\x,\y}{\tc}$ for the two chains to coalesce
starting from $\x$ and $\y$.
If $\x$ and $\y$ have distance $2\ell$,
we denote by $\Ml$  the expected time to reach distance $2\ell-2$. That is,
$$\Ml=\Expec{\x,\y}{\tau_{2\ell-2}}.$$
Similarly, if $\x$ and $\y$ have distance $2\ell-1$,
we denote by $\ml$  the expected time to reach distance $2\ell-2$.
That is,
$$\ml=\Expec{\x,\y}{\tau_{2\ell-2}}.$$
Notice that,
if $\hamming(\x,\y)=\hamming(\x',\y')$ then
$$
	\Expec{\x,\y}{\tau_k}=\Expec{\x',\y'}{\tau_k}
$$
for all $k$, and thus the $\Ml$ and $\ml$ are well defined.

From Lemma~\ref{le:evenl} and Lemma~\ref{le:oddl},
we have the following relations
\begin{eqnarray*}
\Ml&=&1+\Ml\cdot\left(1-\frac{2\ell}{n}\cdot\frac{2}{1+e^\beta}\right)+
	\ml\cdot\frac{2\ell}{n}\cdot\frac{2}{1+e^\beta} \\
\ml&=&1+
	\ml\cdot\frac{n-2\ell+1}{n}\cdot\frac{2}{1+e^\beta}+
	\Ml\cdot\frac{n-2\ell+1}{n}\cdot\left(1-\frac{2}{1+e^\beta}\right).\\
\end{eqnarray*}
Simple algebraic manipulations give
\begin{eqnarray*}
\ml&=&
	\frac{n}{2\ell-1}\left(
		1+\frac{n-2\ell+1}{2\ell}\cdot\frac{e^\beta-1}{2}
			\right)
	\\
\end{eqnarray*}
and
\begin{eqnarray*}
\Ml&=&\ml+\frac{n}{2\ell}\cdot\frac{1+e^\beta}{2}\\
%    &=&\frac{n}{2\ell-1}\left(
% 		1+\frac{n-2\ell+1}{2\ell}\cdot\frac{e^\beta-1}{2}
% 			\right)+
% 	\frac{n}{2\ell}\cdot\frac{1+e^\beta}{2}\\
%    &=&\frac{n}{2\ell-1}+\frac{n}{2\ell}\left(
% 		\frac{n-2\ell+1}{2\ell-1}\cdot\frac{e^\beta-1}{2}+
% 	\frac{1+e^\beta}{2}
% 		\right)\\
%    &=&\frac{n}{2\ell-1}+\frac{n}{2\ell}\left(
% 	\frac{n}{2\ell-1}\cdot\frac{e^\beta-1}{2}-\frac{e^\beta-1}{2}
% 	+\frac{1+e^\beta}{2}
% 		\right)\\
   &=&\frac{n}{2\ell-1}+\frac{n}{2\ell}\left(
	\frac{n}{2\ell-1}\cdot\frac{e^\beta-1}{2}+1
		\right)\\
   &\leq&\frac{n}{2\ell-1}
		\left(
   			\frac{n}{2\ell-1}\cdot\frac{e^\beta-1}{2}+2
		\right)\\
   &\leq&n\left(
		n\cdot\frac{e^\beta-1}{2}+2
		\right)\,.\\
\end{eqnarray*}
Hence,
$$\Expec{\x,\y}{\tc}\leq 1 +
	\sum_{\stackrel{2\leq\ell\leq n}{\ell\textrm{ even }}} \Ml\leq
   \frac{n^2}{2}\left(
		n\cdot\frac{e^\beta-1}{2}+2
		\right) + 1 = \mathcal{O}\left( n^3 e^{\beta} \right) \,.$$
From Markov inequality we have that
$$\Prob{\x,\y}{\tc>t}\leq \frac{\Expec{\x,\y}{\tc}}{t}
$$
and thus, by taking
$t_0=4 \Expec{\x,\y}{\tc}$,
we have $d(t_0)\leq 1/4$.
Therefore, by using Theorem~\ref{thm:coupling}, we have that
 $$\tm = \mathcal{O}\left( n^3 e^{\beta} \right) \,.$$

%% file: trunk/conclusions.tex
In this paper we studied strategic games where at every run a player 
is selected uniformly at random and she is assumed to choose her strategy for 
the next run according to the \emph{logit dynamics}: a noisy best-response dynamics 
where the noise level is tuned by a parameter $\beta$. 
Such dynamics defines a family of ergodic Markov chains, 
indexed by $\beta$, over the set of strategy profiles.

We proposed the stationary distribution of these Markov chains as solution concept for games where players have bounded rationality or limited knowledge about the system. Since this solution concept does not assume full rationality of agents, it avoids one of the main drawbacks of many classical equilibria concepts. Moreover, the stationary distribution of an ergodic Markov chain always exists, it is unique, and the chain converges to such a distribution from any starting state.

In order to evaluate the long-term performance of the system, on the one hand we analyzed the expected social welfare when the strategy profiles are random according to the stationary distribution, on the other hand we studied the \emph{mixing time}, i.e. how long it takes, for a chain starting at an arbitrary profile, to get close to its stationary distribution.

In this paper we applied this approach to some simple but well-studied games with a constant number of players: the \ck\ game, that obtains the worst Price of Anarchy bound between linear congestion games, and the $2 \times 2$ coordination games considered in the seminal paper about logit dynamics~\cite{blumeGEB93}. We also considered two simple $n$-player games, the OR game and the XOR game: the analysis of the mixing time turned out to be far from trivial even for such simple games. The above games highlight a twofold behavior: for some games, namely \ck\ game and OR game, the mixing time can be upper bounded by a function independent of $\beta$, whereas the mixing time for the other games depends exponentially on the noise parameter $\beta$.

%We would like to spot the features of a game that distinguish between this two classes. 

The main goal of our line of research is to investigate logit dynamics for notable classes of $n$-player games.
It would also be interesting to consider variations of the logit dynamics where players update their strategies simultaneously or where the noise is not uniform between players.

We have seen that, for some games and for some values of $\beta$, the mixing time can be exponential in the number of players. When it takes such a long time to reach the stationary distribution, it would be interesting to investigate the evolution of the system in the \emph{transient} phase of the logit dynamics.

%% file: trunk/appendix.tex
\section{Markov Chains' Summary}
\label{apx:markov}
We summarize the main tools we use to bound the mixing time of Markov chains (for a complete description of such tools see, for example, Chapters~7.2, 12.2 and 14.2 of~\cite{lpwAMS08}).

\smallskip \noindent
For an irreducible and aperiodic Markov chain $\mathcal{M}$ with finite state space $\Omega$, transition matrix $P$, and stationary distribution $\pi$, the \emph{mixing time} is defined as
$$
t_{\text{mix}}(\varepsilon) = \min\left\{ t \in \mathbb{N} \; : \; \max \{ \| P^t(x, \cdot) - \pi \|_{\text{TV}} \,:\, x \in \Omega \} \leqslant \epsilon \right\}
$$
where 
$P^t(x, \cdot)$ is the distribution at time $t$ of the chain starting at $x$, 
$\| P^t(x, \cdot) - \pi \|_{\text{TV}} = \frac{1}{2} \sum_{y \in \Omega}| P^t(x, y) - \pi(y)|$ 
is the \emph{total variation distance}, 
and $\varepsilon > 0$ is a constant smaller than $1/2$.
For convenience, it is usually set to $\varepsilon = 1/4$ or $\varepsilon = 1/2e$. 
If not explicitly specified, when we write $t_{\text{mix}}$ we mean $t_{\text{mix}}(1/4)$.

\begin{theorem}[Path coupling]\label{theorem:pathcoupling}
Let $\mathcal{M} = \{ X_t \,:\, t \in \mathbb{N} \}$ be an irreducible and aperiodic Markov chain with finite state space $\Omega$ and transition matrix $P$. Let $G = (\Omega, E)$ be a connected graph, let $\ell\,:\, E \rightarrow \mathbb{R}$ be a function assign \emph{weights} to edges such that $\ell(e) \geqslant 1$ for every edge $e \in E$, and let $\rho\,:\, \Omega \times \Omega \rightarrow \mathbb{R}$ be the corresponding path distance, i.e. $\rho(x,y)$ is the length of the (weighted) shortest path in $G$ between $x$ and $y$.

\noindent Suppose that for every edge $\{ x,y \} \in E$ a coupling $(X,Y)$ of distributions $P(x, \cdot)$ and $P(y, \cdot)$ exists such that $\Expec{x,y}{\rho(X,Y)} \leqslant \ell(\{x,y\}) e^{-\alpha}$ for some $\alpha > 0$. Then the mixing time $t_{\text{mix}}(\varepsilon)$ of $\mathcal{M}$ is
$$
t_{\text{mix}}(\varepsilon) \leqslant \frac{\log(\text{diam}(G)) + \log(1/\varepsilon)}{\alpha}
$$
where $\text{diam}(G)$ is the (weighted) diameter of $G$.
\end{theorem}

\begin{theorem}[Bottleneck ratio]\label{theorem:bottleneck}
Let $\mathcal{M} = \{ X_t \,:\, t \in \mathbb{N} \}$ be an irreducible and aperiodic Markov 
chain with finite state space $\Omega$, 
transition matrix $P$ and stationary distribution $\pi$. 
Let $S \subseteq \Omega$ be any set with $\pi(S) \leqslant 1/2$. 
Then the mixing time is
$$
t_{\text{mix}}(\varepsilon) \geqslant \frac{1-2\epsilon}{2 \Phi(S)}
$$
where
$$
\Phi(S) = \frac{Q(S,\overline{S})}{\pi(S)} \quad \mbox{ and } \quad Q(S, \overline{S}) = \sum_{x \in S, \, y \in \overline{S}} \pi(x) P(x,y).
$$
\end{theorem}

For a reversible transition matrix $P$ of a Markov chain with finite state space
$\Omega$, the {\em relaxation time} $t_{\text{rel}}$ is defined as
$$
t_{\text{rel}}={\frac{1}{1-\lambda^\star}}
$$
where $\lambda^\star$ is the largest absolute value of an eigenvalue 
other than $1$,
$$
\lambda^\star = \max\{ |\lambda| \;:\; \lambda \text{ is an eigenvalue of } P, \, \lambda \neq 1 \} \,.
$$
Notice that all the eigenvalues have absolute value at most $1$, $\lambda = 1$ is an eigenvalue, and for irreducible and aperiodic chains, $-1$ is not an eigenvalue. Hence $t_{\text{rel}}$ is positive and finite.

We have the following theorem.
\begin{theorem}[Relaxation time]\label{theorem:relaxation}
Let $P$ be the transition matrix of a reversible, irreducible, and aperiodic Markov chain with state space 
$\Omega$ and stationary distribution $\pi$. Then 
$$
(t_{\text{rel}}-1)\log\left({\frac{1}{2\epsilon}}\right)
\leq t_{\text{mix}}(\epsilon)\leq
\log\left({\frac{1}{\epsilon\pi_{\text{min}}}}\right) t_{\text{rel}}
$$
where 
$\pi_{\text{min}}=\min_{x\in\Omega} \pi(x)$.
\end{theorem}